\documentclass[preprint]{aastex}

\shorttitle{A Photometric Study of NGC 5822}
\shortauthors{Carraro, Anthony-Twarog, Costa, Jones, Twarog}

\begin{document}


\title{A $UBVI$ and $uvbyCa$H$\beta$ Analysis of the Intermediate-Age Open Cluster, NGC 5822}


\author{Giovanni Carraro\altaffilmark{1}}
\affil{ESO, Alonso de Cordova 3107, Santiago de Chile, Chile}
\email{gcarraro@eso.org}

\author{Barbara J. Anthony-Twarog}
\affil{Department of Physics and Astronomy, University of Kansas, Lawrence, KS 66045-7582, USA}
\email{bjat@ku.edu}

\author{Edgardo Costa}
\affil{Departamento de Astronomía, Universidad de Chile, Casilla 36-D, Santiago, Chile}
\email{costa@das.uchile.cl}

\author{Bryce J. Jones}
\affil{Department of Physics and Astronomy, University of Kansas, Lawrence, KS 66045-7582, USA}
\email{bjj8383@ku.edu}

\and

\author{Bruce A. Twarog}
\affil{Department of Physics and Astronomy, University of Kansas, Lawrence, KS 66045-7582, USA}
\email{btwarog@ku.edu}



\altaffiltext{1}{On leave from Dipartimento di Astronomia, Universit\'a di Padova,
Vicolo Osservatorio 3, I-35122, Padova, Italy}


\begin{abstract}
NGC 5822 is a richly populated, moderately nearby, intermediate-age open cluster covering an area larger than
the full moon on the sky. A CCD survey of the cluster on the $UBVI$ and $uvbyCa$H$\beta$ systems shows that
the cluster is superposed upon a heavily reddened field of background stars with E$(B-V) > 0.35$ mag, while
the cluster has small and uniform reddening at E$(b-y)$ = 0.075 $\pm$ 0.008 mag or E$(B-V)$ = 0.103 $\pm$
0.011 mag, based upon 48 and 61 probable A and F dwarf single-star members, respectively. The errors quoted include both
internal photometric precision and external photometric uncertainties. The metallicity derived from 61 probable 
single F-star members is [Fe/H] = -0.058 $\pm$ 0.027 (sem) from $m_1$
and 0.010 $\pm$ 0.020 (sem) from $hk$, for 
a weighted average of [Fe/H] = -0.019 $\pm$ 0.023, where the errors refer to the internal errors from the photometry alone.
With reddening and metallicity fixed, the cluster age and apparent distance modulus are obtained through a comparison
to appropriate isochrones in both $VI$ and $BV$, producing 0.9 $\pm$ 0.1 Gyr and 9.85 $\pm$ 0.15, respectively.
The giant branch remains dominated by two distinct clumps of stars, though the brighter clump seems a better match to
the core-He-burning phase while the fainter clump straddles the first-ascent red giant branch. Four potential new clump
members have been identified, equally split between the two groups. Reanalysis of the $UBV$ two-color data extending
well down the main sequence shows it to be optimally matched by reddening near $E(B-V)$ = 0.10 rather than the older
value of 0.15, leading to [Fe/H] between -0.16 and 0.00 from the ultraviolet excess of the unevolved dwarfs. The
impact of the lower reddening and younger age of the cluster on previous analyses of the cluster is discussed. 

\end{abstract}



\section{Introduction}
The canonical rationale for the study of star clusters is their value as star systems of homogeneous age, 
distance, and composition, making them ideal testing grounds for stellar structure and evolution, as well 
as probes of Galactic evolution. However, the growing evidence from Milky Way globular clusters \citep{pi09} and 
Magellanic Cloud open clusters \citep{ru10, gou11} casts serious doubt on the degree of temporal and chemical 
homogeneity for many, if not all, dense, richly populated clusters more than $10^8$ years old. Moreover, 
for nearby clusters the cluster diameter can be comparable in scale to the cluster distance, making isolation 
of the cluster from the local field star population a challenge in the absence of radial-velocity and/or
proper-motion membership, a particular problem for the lower-luminosity stars. For more 
distant clusters, variable reddening and field star contamination can impact the interpretation 
of observations of a majority of the members of the cluster. Studies of individual clusters therefore are 
often driven by a specific aspect of stellar or galactic evolution that can best be probed due to the correct 
combination of cluster parameters and the extent to which they are reliably known, making some clusters iconic 
examples of a given class, as in M67 and NGC 188 for old open clusters of solar composition and the Hyades for 
metal-rich clusters of intermediate age.

The focus of this investigation, NGC 5822, is no exception. Some basic properties of the cluster are summarized
in Table 1, emphasing the most recent determinations of metallicity, reddening, distance, and age from a
variety of techniques. With an age of $\sim$1 Gyr and a metallicity near 
solar, it exemplifies the properties typical for star clusters of its age within 
1 kpc of the sun \citep{tat97}. In the context of Galactic evolution studies, its role is limited to that of a 
consistency check for the general cluster population near the sun. For stellar evolution, however, it falls 
at or near an age range when important transitions are occurring among the stars dominating the brighter 
end of the color-magnitude diagram (CMD) and many less populous star clusters are evaporating 
into the field star background \citep{jtl88,jp94,fa06,pis06,ros10}.

For example, stars leaving the NGC 5822 main sequence have masses approaching the range where He-ignition switches 
from quiescent to non-quiescent under degenerate conditions, leading to potentially significant changes in 
the distribution of stars in the red giant region \citep{gir98, gir99}. The CMD just below the turnoff is 
populated by stars with masses where convection first appears as a function of decreasing main sequence mass, 
possibly producing the poorly understood B{\"o}hm-Vitense gap \citep{bc74}. Slightly fainter, one expects a 
fully-formed, though potentially deeper, Li-dip than found in the slightly younger Hyades, but less distorted 
by post-main-sequence evolution than the equivalent feature seen in slightly older clusters like NGC 3680, NGC 752, and IC 
4651 \citep{pas04, att04, att09}. The derived cluster age places it on the young side of the Vaughan-Preston gap, but
how rapidly the chromospheric activity declines across the gap depends to a significant degree on exactly how young 
the cluster is \citep{pac09}.  

Of primary importance, however, is the combination of a modest distance modulus ($\mu \equiv (m-M) \leq 10$) and a rich 
stellar population. Early CCD \citep{wat91} and photographic work \citep{tam93} demonstrated that, counting 
stars within a $15\arcmin$ radius of the cluster, the CMD isn't significantly impacted by field star contamination down to 
at least two magnitudes below the turnoff, despite its position in the galactic plane in the 
general direction of the galactic center. Moreover, the giant branch and 
unevolved main sequence are exceptionally rich compared to other nearby clusters of equal or smaller 
age \citep{rc00}. However, in the absence of radial-velocity measures for stars below the turnoff and 
proper-motion membership estimates for any stars in the field, detailed spectroscopic study of the fainter stars on the 
unevolved main sequence would be subject to non-negligible contamination by possible background 
stars, negatively impacting the efficiency of large telescope time devoted to the cluster. With an interest 
in a comprehensive spectroscopic attack on the cluster members from the giant branch to the unevolved 
main sequence below solar-mass stars in mind, it was decided that a multiband photometric survey of the cluster 
could prove valuable given the right combination of broad-band and intermediate-band filters. As 
demonstrated by the results of this investigation, $UBVI$ CCD photometry meshed with intermediate-band 
extended Stromgren indices, $uvbyCa$H$\beta$, allows easy isolation of the cluster members of NGC 5822 
from the rich background contamination, while supplying precise estimates of the key cluster parameters 
of reddening, metallicity, age, and distance.

The outline of the paper is as follows: Sec. 2 discusses the CCD observations and their reduction to the 
standard system for both broad-band and intermediate-band photometry; Sec. 3 demonstrates the use of the 
intermediate-band and narrow-band data to isolate the nearby cluster from the more heavily reddened 
background stars and to precisely define the cluster metallicity and reddening. Highly probable single 
cluster members from the field are used in Sec. 4 to define the cluster age and distance. In Sec. 5, the adopted 
members are used to isolate the cluster in a broad-band color-color diagram and to rederive the metallicity from 
the $UBV$ photometry of the dwarfs while Sec. 6 contains a summary of our conclusions.

\section{Observations and Data Reduction}
\subsection{$UBVI$ Photometry}       

For $UBVI$ data, NGC 5822 was observed during two runs; Table 2 provides a log of the $UBVI$ observations.
During the first run on March 19, 2009, only the central field (E) was observed, while
over the second run from March 11-14, 2010, an additional 4 pointings (A, B, C and D) were observed, leading
to an areal coverage 40 arcmin on a side (Fig.1). The observations were carried out with the Y4KCAM camera 
attached to the Cerro Tololo Inter-American Observatory (CTIO) 1-m telescope, operated by the SMARTS 
consortium.\footnote{\tt http://http://www.astro.yale.edu/smarts} This camera is equipped with an STA
4064$\times$4064 CCD\footnote{\texttt{http://www.astronomy.ohio-state.edu/Y4KCam/detector.html}}
with 15-$\mu$m pixels, yielding a scale of 0.289 \arcsec /pixel and a field-of-view (FOV) of 
20\arcmin $\times$ 20\arcmin\ at the Cassegrain focus of the CTIO 1-m telescope. The CCD was operated 
without binning at a nominal gain of 1.44 e$^-$/ADU, implying a readout noise of 7 e$^-$ per quadrant, with four 
different amplifiers supplying the simultaneous readout.

All observations were carried out under photometric conditions with seeing below 1.2\arcsec.
Our $UBVI$ instrumental photometric system was defined using standard broad-band Kitt Peak $UBVI_{kc}$ 
filters.\footnote{\texttt{http://www.astronomy.ohio-state.edu/Y4KCam/filters.html}}
To define the transformation to the standard Johnson-Kron-Cousins system and to correct for atmospheric extinction, 
standard stars were observed in Landolt fields PG-1047 and SA-98 \citep{lan92} multiple times at air masses 
ranging from $\sim1.05$ to $\sim2.4$, and covering a color range of -0.3 $\leq (B-V) \leq$ 1.7 mag.

Basic calibration of the CCD frames was done using the Yale/SMARTS y4k reduction script
based on the IRAF\footnote{IRAF is distributed by the National Optical Astronomy Observatory, which 
is operated by the Association of Universities for Research in Astronomy, Inc., under cooperative agreement with
the National Science Foundation.} package CCDRED. For this purpose, bias frames and twilight sky flats were 
taken every night.  Photometry was then performed using the IRAF DAOPHOT and PHOTCAL packages. Instrumental 
magnitudes were extracted following the point-spread function (PSF) method \citep{stet87}. A quadratic, spatially
variable, master PSF (PENNY function) was adopted. Aperture corrections were determined using aperture photometry 
of typically 10 to 20 bright, isolated stars in the field; corrections were found to vary from 0.160 to 0.290 mag, 
depending on the filter. The PSF photometry was then aperture-corrected, filter by filter.

After removing problematic stars and stars having only a few observations in the \citet{lan92} catalog, the 
photometric transformation solutions from 272 measurements per filter were found to be:

\noindent
$ U = u + (3.099\pm0.010) + (0.45\pm0.01) \times X - (0.008\pm0.006) \times (U-B)$ \\
$ B = b + (1.951\pm0.012) + (0.27\pm0.01) \times X - (0.141\pm0.007) \times (B-V)$ \\
$ V = v + (1.892\pm0.007) + (0.15\pm0.01) \times X + (0.031\pm0.007) \times (B-V)$ \\
$ I = i + (2.696\pm0.011) + (0.08\pm0.01) \times X + (0.016\pm0.008) \times (V-I)$ \\

The final {\it rms} residuals of the fit to the observations were 0.030, 0.015, 0.010, and 0.010 mag in $U$, $B$, $V$
and $I$, respectively.

Global photometric errors were estimated using the scheme discussed in Appendix A1 of \citet{pc01}, which takes 
into account the errors resulting from the PSF fitting procedure within ALLSTAR, and the calibration 
errors corresponding to the zero point, color term, and extinction errors. In Fig. 2 we present our global 
photometric errors in $V$, $(B-V)$, $(U-B)$, and $(V-I)$ plotted as a function of $V$. For clarity, the data have 
been sorted into an inner region defined by field E (open circles) and an outer region containing all stars not 
within E (crosses). The bin centers for $V$ have been offset for the outer sample versus the inner region 
for easier comparison. Quick inspection shows that stars brighter than $V \sim 21$ have average errors in $V$ lower 
than 0.05 mag, while the average errors remain lower than 0.10 mag in $(B-V)$ for $V$ brighter than 20 and 18.5
in the inner and outer regions, respectively. For $(V-I)$, the average errors remain below 0.10 for $V$ brighter
than 21.75 and 21.00, respectively. Predictably higher errors are seen in $(U-B)$, with the error average of 0.1 mag
occurring at $V$ = 18.5 (inner) and 17.5 (outer), respectively.

\subsection{Completeness and Astrometry}

Completeness corrections were determined by running artificial star experiments on the data. Basically, several 
artificial frames were created by adding artificial stars to the original frames. These stars were added 
at random positions and had the same color and luminosity distribution as the true sample. To avoid generating 
overcrowding issues, in each experiment artificial stars were added only at a rate up to 20\% of the original 
number of stars. Depending on the frame, this meant that between 1000 and 5000 stars were added. Reprocessing
and analysis of the artificial frames leads to the conclusion that the completeness level of our $UBVI$ photometry 
is better than 90\% down to $V = 19.5$.

Our optical catalogue was cross-correlated with 2MASS and CCD pixel coordinates were converted to RA and DEC for J2000.0 
equinox, thus providing 2MASS-based astrometry. Our final $UBVI$ photometric catalog for all stars 
down to $V\sim$21 mag for any index where the final error is less than $\pm$0.10 mag is listed in Table 3.
Stars are identified through their RA and DEC on the 2MASS system.

\subsection{Comparison with Previous Photometry}

The largest broad-band photometric survey of NGC 5822 to date has been that of \citet{tam93}, which includes 
photoelectric data on the $UBV$ system for 144 stars, though not all of the stars have $U$ data. These photoelectric
standards were used to calibrate 8 photographic plates each in $B$ and $V$, leading to precise photographic magnitudes and colors
for over 600 stars in an area of 15\arcmin\ radius centered on the cluster. Of the 144 
photoelectric standards, 135 between $V$ = 9.49 and 15.27 were found to overlap with the observations 
in Table 3. The 9 stars without matches, i.e. without CCD data, include 
a number of stars that were too bright to avoid saturation on the CCD images, as well as a few that were probable 
composites. One star, 114 on the identification system from WEBDA, showed significant residuals in all 
magnitudes and indices and was dropped from the comparison. The average residuals, in the sense (TAM - Table 3), 
and their standard deviations are 0.003 $\pm$0.034 and 0.007 $\pm$ 0.034 for $V$ and $B-V$, respectively. 
From 127 stars with $UBV$ data, the analogous residual is $-0.015 \pm$ 0.053 for $U-B$; if the 9 stars with 
absolute residuals greater than 0.10 mag are excluded, the offset becomes $-0.010 \pm$ 0.038 mag. The trends 
among the residuals for the various indices as a function of magnitude and color are shown in Fig. 3.  

A more comprehensive comparison is available through the photographic data which, due to the significant number of
plates, has internal precision competitive with the single-measurement photoelectric data within the cluster. The residuals
in $V$ and $B-V$, in the sense (Table 3 - PG), as a function of $V$ are illustrated in Fig. 4 for 555 stars that 
overlap with the CCD sample of Table 3. The mean offset in $V$ is 0.025 $\pm$ 0.046 mag, implying that the photographic
magnitudes are too bright, with a slight trend of an increasing offset at fainter $V$. The explanation for this is
straightforward in that the photographic data are based upon iris astrophotometer measures which cannot as easily exclude faint
companions from the iris measurement as PSF fitting. Thus, stars with faint optical companions 
will appear brighter in the photographic sample and the impact will be larger for less luminous stars. Support for 
this explanation comes from the asymmetry in the residuals at a given magnitude, with more scatter on the 
positive $\Delta$$V$ side, and the fact that the scatter in the residuals for $B-V$ is typically the same or smaller 
than that for the residuals in $V$ at a given magnitude. Since $B$ and $V$ are calibrated independently, this 
lack of growth in the scatter for the combined magnitudes can only occur if the errors in $B$ and $V$ are 
correlated, i.e. the faint companion brightens the magnitude determination in both filters.  

The obvious trend in the residuals in $B-V$ can also be attributed to the nature of the photographic calibration, particularly
in $B$. As shown in Fig. 3, of 144 photoelectric standards only 9 are fainter than $V$ = 14 and only 2 of these
fall below $V$ = 14.75. Nonlinear terms used in the calibration of the photographic magnitudes to account for the
plate response at the faintest magnitudes were very sensitive to these last few standards and modest errors in 
these standards could easily produce trends of the type illustrated in Fig. 4. A similar effect can be seen in a
comparable discussion by \citet{nor96} of the similarly processed $V$ photographic data for NGC 3680 \citep{att91}. 
It should be noted that the probable existence of an offset approaching 0.1 mag in the photographic $B-V$ values near
$V$ = 14.5 was predicted by \citet{pac10} based upon high-dispersion spectroscopic determinations of the
temperatures of two solar-type dwarfs.  

\subsection{$uvbyCa$H$\beta$ Photometry}
Intermediate and narrow-band imaging of NGC 5822 was accomplished over five nights in June 2010 using the CTIO 0.9-m 
telescope, operated by the SMARTS consortium. The telescope is equipped with a TEK $2048 \times 2048$ CCD imager at the $f$/13.5
Cassegrain focus, producing a scale of 0.396 \arcsec /pixel and a FOV of 13.5\arcmin\ $\times$ 13.5\arcmin. The images 
were taken unbinned and read out in QUAD mode at a selected gain of 1.52 e$^-$/ADU. All seven filters were 
3 $\times$ 3 inches and came from a set owned jointly by the University of Kansas and Mt. Laguna Observatory.

The cluster was imaged every night, though only four of the five nights were photometric and not every filter was
included in every night's observations. Three overlapping fields in the north-south direction were studied, leading to
final areal coverage of approximately 13.5\arcmin\ $\times$ 25\arcmin, overlapping nicely with region E of Fig. 1. Exposure
times in all filters were staggered in length to supply comparable precision from the brightest stars in the field
to the unevolved main sequence below $V$ = 16.5.

The first four nights of the five-night run were photometric. On two of the photometric nights, a combination of field star 
and cluster standards was observed to establish transformation equations for $V$, $(b-y)$, $hk$, $m_1$ and $c_1$; 
an alternate pair of photometric nights was used to calibrate the first three indices plus H$\beta$.  All standards and program 
stars were utilized for extinction correction determinations if their observations spanned a suitable range of airmass.  
A common set of extinction correction coefficients was determined and used for all four photometric nights.

Standard IRAF routines were used to perform initial processing of the frames, i.e. bias-subtraction and flat-fielding, using 
dome flats for the $y$ frames and sky flats for the other six filters. A fairly comprehensive discussion of the
procedure for obtaining PSF-based instrumental magnitudes and merging multiple frames of a given filter can be
found in \citet{att00}.

Our calibrations to the standard extended Str\"omgren system are based on aperture photometry in the program cluster, in 
three open clusters (NGC 6633, NGC 3680 and IC 4651), and of field star standards on each photometric night.  For all 
four photometric nights the seeing was 1.4\arcsec\  to 1.8\arcsec, permitting the use of apertures 4.8\arcsec\ in radius, surrounded 
by a sky annulus of comparable area. A number of sources were consulted for field star standard values, including the catalog 
of \citet{tat95} for $V$, $b-y$ and $hk$ indices, catalogs of $uvby$H$\beta$ observations by 
\citet{ols83,ols93,ols94}, and compilations of H$\beta$ indices by \citet{hm98} and \citet{sn89}. 
For stars in open clusters, targeted used was made of $uvby$H$\beta$ photoelectric standards in NGC 6633 \citep{ed76}, 
red giants in NGC 3680 for which $V$, $b-y$ and $hk$ photoelectric photometry exists \citep{att04}, and 
$uvby$ photometry of red giants in IC 4651 \citep{att00}.

Following standard procedures for Str\"omgren photometry, a single $(b-y)$ calibration equation was derived for 
warmer dwarfs and giant stars; too few standards were available for cooler dwarfs to determine an independent 
calibration for dwarfs with $(b-y)\geq 0.42$ so, while the internal precision may be quite good, $m_1$ and 
especially $c_1$ indices for cooler dwarfs may be subject to color-dependent offsets relative to standard relations,
a point we will return to in Sec. 3. Calibrations of $m_1$ and $c_1$ for cooler giants are determined 
independently from calibrations applied to bluer dwarfs.  All photoelectric standards, field stars and cluster stars 
alike, were used to determine slopes and color terms for the calibration equations, summarized in Table
 4. An 
independent zero point was determined for each calibration equation on each night, based preferentially on field 
star standards augmented by photoeletric photometry in the three open clusters.  

The extension of calibration equations to the merged profile-fit photometry for stars in NGC 5822 is 
facilitated by determining the average differences between profile-fit indices and indices determined 
from aperture photometry in the cluster on each photometric night. It was possible to determine the average 
difference between the aperture and profile-fit indices for 50 to 90 stars in NGC 5822 with a standard deviation 
of 0.03 mag or less, so that the ``aperture corrections" for each index could be determined to very high precision. 
In this manner, the calibration equations from each photometric night may be transformed to the aperture photometry 
in the program cluster, and then by extension to the profile-fit indices in NGC 5822, with an independent zeropoint 
for each equation from each photometric night. The precision of the combined zeropoint applied to the NGC 5822 
photometry is also indicated in Table 4. Contributions to this quoted error arise from the standard errors of the mean
({\it sem}) from each aperture correction and the {\it sem} of the zeropoint from the calibration equation.   

Final photometry on the $uvbyCa$H$\beta$ system can be found in Table 5. For consistency with Table 3,
(X,Y) positions are presented as right ascension and declination on the 2MASS system. A plot of the {\it sem} for each 
index as a function of $V$ can be seen in Fig. 5. Standard errors of the mean for the indices 
are calculated by combining the errors for individual filters in quadrature and are defined solely by the 
internal, i.e. frame-to frame, precision of the of the individual filters. Because of the large number of 
frames, the {\it sem} remain quite small to $V \geq 16.5$.

To provide additional validation of our calibration methodology through aperture photometry on photometric 
nights, we applied identical precepts to observations in NGC 3680 and IC 4651, treating each cluster as 
a program object.  An {\it ex post facto} comparison of our calibrated photometry in NGC 3680 to 
photoelectric $V$, $b-y$ and $hk$ indices for red giants \citep{att04} demonstrates excellent agreement with 
the 2004 photometry.  Mean differences, in the sense (PE - CCD), for 26 stars for $V$, $b-y$ and $hk$ are $0.004 \pm 0.006$, 
$0.003 \pm 0.008$ and $-0.016 \pm 0.015$ respectively, where the indicated errors are standard deviations. From 
45 dwarfs brighter than $V$ = 14, the comparable offsets relative to the 2004 CCD data in NGC 3680, in the sense (2004 - Table 5), 
are 0.004 $\pm$ 0.006, 0.003 $\pm$ 0.009, 0.005 $\pm$ 0.012, 0.006 $\pm$ 0.012, -0.013 $\pm$ 0.018, and 0.005 $\pm$ 0.007
in $V$, $b-y$, $m_1$, $c_1$, $hk$, and H$\beta$, respectively. 

Similar comparisons were constructed between calibrated indices for IC 4651 stars to $uvby$ indices for 
red giants \citep{att04} as well as turnoff and main sequence stars \citep{att87,meibom}.  An additional 
comparison of H$\beta$ indices was possible with respect to indices from \citep{att87}.  For 14 red giants, 
the mean differences between indices from \citet{att87} and calibrated indices from this run are $-0.020 \pm 0.037$, 
$0.016 \pm 0.021$, $-0.010 \pm 0.031$, $0.029 \pm 0.032$, and $-0.001$ (1 star only) for $V$, $b-y$, $m_1$, $c_1$ and H$\beta$ 
respectively.  From comparisons with main sequence and turnoff stars \citep{att87}, the 
average differences are $-0.021 \pm 0.026$, $0.011 \pm 0.009$, $0.017 \pm 0.014$, $0.037 \pm 0.020$, $-0.004 \pm 0.012$ 
for the same indices. No H$\beta$ photometry was included in \citet{meibom} and it is likely that the 
calibration precepts for red giants are substantially different, so the following comparisons, 
in the sense (MEI-this paper), are derived using 150 main sequence stars between $V=11.5$ and $14.0$: 
$-0.030 \pm 0.013$, $0.007 \pm 0.027$, $-0.003 \pm 0.027$ and $0.040 \pm 0.026$ for $V$, $b-y$, $m_1$ and $c_1$.

\subsection{Comparison to Previous Photometry} 
On the $uvby$H$\beta$ system, photoeletric data are available for 21 stars from the survey by \citet{stet81}. 
For the 14 stars in common to the two studies, the mean offsets, in the
sense (Table 5 - ST), are $-0.013 \pm 0.011$, $0.011 \pm 0.025$, $0.016 \pm 0.040$, $0.029 \pm 0.043$, and $0.000 \pm 0.020$ 
for $V$, $b-y$, $m_1$, $c_1$, and H$\beta$, respectively, where the indicated errors are standard deviations. 

On a much larger scale, we can also compare the $V$ system from the $y$ filter with that tied to the traditional
broad-band $V$ of Table 3. For just under 2250 stars to $V$ = 18 that overlap between Table 3 and Table 5, the
mean offset, in the sense (Table 3 - Table 5), is 0.000 $\pm$ 0.033. From Fig. 6, where the residuals in $V$ are
plotted as a function of $V$, it should be emphasized that among the 67 stars brighter than $V$ = 12.5, the
average offset is closer to -0.015 $\pm$ 0.021. The reason for this slight discontinuity remains unknown, but it has no
impact on our conclusions.

\section{The CMD and Cluster Parameters: Membership, Reddening and Metallicity}
The traditional CMD based upon ($V, B-V$) for all stars within the CCD fields is shown in Fig. 7. The key structures of the CMD
are apparent: a richly populated red giant branch with what appear to be two concentrations of stars with
$B-V$ = 0.96 to 1.05 at $V$ = 10.35 and $B-V$ = 0.99 to 1.05 at $V$ = 10.8, a well-defined Hertzsprung gap
with the turnoff separated from the giant branch in color by an amount typical of clusters with an age of $\sim$1 Gyr, as
well as a handful of blue stragglers. The main sequence is richly populated with a sharp blue edge but
significant scatter to the red side. The impact of the field star population rises sharply at all colors for 
$V$ fainter than 13 and grows with increasing $V$. The role of the probable field star contamination
in the outer regions of the cluster becomes even
more apparent when one restricts the sample to the core region, i.e. stars within a rectangular zone $13.5^{\prime}$
on a side centered on the cluster and chosen to overlap with the central field of the intermediate-band images.
This CMD is shown in Fig. 8. In addition to the obvious reduction in the number of points, the main sequence and
turnoff region are much better defined with significantly less scatter. Few of the 
stars with $V \leq 13$ are likely to be non-members. Note that part of the improved cluster delineation is a 
byproduct of the more precise photometry
in the cluster core region due to the repeated observations of region E over two observing runs while stars outside
E were observed during only one run. Still, contamination by field stars below $V$ = 13 is non-negligible even in the
core.

For comparison, we can turn to the ($V, b-y$) based CMD for all stars in Table 5 brighter than $V$ = 18, as
shown in Fig. 9. Stars with {\it sem} errors in $b-y$ below 0.015 are shown as open circles, while stars with errors greater than 
this cutoff to an {\it sem} limit of 0.15 mag are plotted as crosses. Unlike Fig. 7, these data come only from a strip $13.5^{\prime}$ 
by 25$^{\prime}$ centered on the cluster core. The turnoff region and giant branch are again richly defined, while contamination 
by the field sample kicks in for $V$ fainter than 13. The less populated red giant branch makes definitive identification of the two potential 
clumps at $V$ = 10.35 and 10.8 ($b-y$ = 0.62 to 0.68) less obvious. Finally, Fig. 10 shows the CMD for the same region as Fig. 8. 
The reduction in field star contamination is apparent, though not as dramatic as the comparison of Fig. 7 to Fig. 8 due to the 
smaller differential between the areas.

While restricting our discussion to core stars would reduce the impact of field star contamination on the
cluster parameters, it still remains a growing issue for stars fainter than $V$ = 13 on the main sequence
because this magnitude-color range is occupied by the F dwarfs which are used to define the cluster reddening and
metallicity. One can, however, appeal to the reddening and metallicity-independent H$\beta$ indices to 
enhance the separation between members and non-members. Fig. 11 demonstrates the principle. Stars in Fig. 10 
with at least 2 observations each in H$\beta$ wide and narrow, an {\it sem} below 0.015 mag in $b-y$ and 0.012 mag 
for the H$\beta$ index are plotted in Fig. 11. Filled circles are the
10 giants brighter than $V$ = 11.05. The trend in Fig. 11 is gratifyingly obvious. The tight band of stars
at the bluest $b-y$ for a given H$\beta$ is defined by cluster members. Moving vertically in the plot, one encounters
a large gap of typically 0.25 mag in $b-y$, followed by a wider band of points with the same two-color profile
as the cluster members. Since reddening can only move stars vertically in the two-color plot, this implies that the
field stars in the CMD are dominated by background stars at a large enough distance that the line of sight 
reddening is at least 0.25 mag larger in $b-y$, or about 0.35 mag larger in E$(B-V)$, than the nearby cluster.  

The tight trend defined by the probable members allows us to isolate other likely members from a 
wider area than the core. A mean relation, shown as a solid line in Fig. 11, was derived for the probable
cluster stars, excluding the giants, in Fig. 11. All stars with $uvby$H$\beta$ photometry, 
$V \leq 17.5$, with at least 2 observations in $b$, $y$ and both $\beta$ filters, errors in $b-y$ and H$\beta$ below 0.015 mag, and
located within 0.04 mag of the solid line in Fig. 11 were selected from Table 5. The CMD for these 201 stars is shown in Fig. 12.
The cluster main sequence and turnoff region are easily identified while the scatter of stars redward of the cluster CMD is 
dramatically reduced to stars occupying the predicted region for binaries (crosses) and a handful of stars that are too bright
relative to the single-star main sequence to be binary members of the cluster (filled circles). The fact that the latter group
falls within the restricted zone around the mean relation in Fig. 11 ensures that they must have low reddening similar to the
cluster, implying that most must be foreground dwarfs. This can be tested using the LC parameter as
derived in \citet{tvat07}. The LC parameter is based upon a combination of the $b-y$, $m_1$, and $c_1$ indices which
easily separates cool dwarfs from giants. For the ten giants in Fig. 11, the LC parameter classifies all ten as evolved stars.
By contrast, for the six stars with $(b-y)_0$ greater than 0.5 among the filled circles of Fig. 12, the LC
indicates that all six are dwarfs. 

An additional photometric check on our sample selection is
provided in Fig. 13 through the $(V, hk)$ CMD. Symbols have the same meaning as in Fig. 12. The value of the $hk$ index is
its weak sensitivity to reddening but strong sensitivity to temperature changes. Every star that deviates significantly from
the mean trend for the cluster (open circles) was identified as a potential binary or non-member in Fig. 12, ensuring that
at their observed $V$ magnitude, these stars are too red for reasons that cannot be explained through reddening or photometric
errors.

Finally, while we have used photometric criteria to identify probable NGC 5822 members (open circles as single stars and crosses as
probable binaries in Figs. 12 and 13) and non-members (all other stars for which $uvbyCa$H$\beta$ photometry is available),
the relative proximity of the cluster might permit us to test the membership
results using absolute proper motions from UCAC3 \citep{zac00}. UCAC3 is a valuable tool for stars down to 
R $\sim$ 16.0, which, in the case of NGC 5822, allows us to probe stars about 5 mag below the turnoff point.
A cross correlation with the UCAC3 database yields 136 probable photometric members and 322 probable photometric non-members.

The resulting vector point diagram is shown in Fig. 14, where probable photometric members are indicated by black
symbols and non-members with blue symbols. The dashed lines indicate the mean values in the proper-motion components. The mean proper-motion components 
($\mu_{\alpha} cos(\delta), \mu_{\delta}$) for the non-member sample are $-3.98 \pm 0.80$ and $-3.29 \pm 0.59$ mas/yr. 
For the 136 photometric members, the comparable numbers are 
 $-7.57 \pm 1.45$ and $-5.32 \pm 1.24$ mas/yr.    

However, most of the scatter in the mean motion among the smaller sample of photometric members is caused by a 
handful of exceptional outliers in the proper-motion plot. If the eight most deviant points are excluded, the resulting 
proper motion components ($\mu_{\alpha} cos(\delta)$ and  $\mu_{\delta}$) become $-7.90 \pm 0.45$ and $-5.82 \pm 0.64$ mas/yr.

The two distributions are different at a significant level, strengthening the case for the photometric classification, though the 
dispersion in both cases is dominated by the individual errors in the astrometric measures.

With the restricted sample of 153 probable single-star photometric members of NGC 5822 (open circles in Figs. 12 and 13), 
we can now derive the cluster reddening. As in past
cluster analyses, use is made of two intrinsic H$\beta$-$(b-y)_0$ relations to define the intrinsic colors. The first from
\citet{ols88} applies to F stars in the H$\beta$ range from 2.58 to 2.72. Additional restrictions on the photometry required
at least two observations in every filter used in $m_1$ and $c_1$, errors in $b-y$, $m_1$, $c_1$, $hk$, and H$\beta$ less than
or equal to 0.015, 0.040, 0.050, 0.050, 0.015, respectively, and $V$ brighter than 17.5. From 61 F dwarfs that meet all
the criteria, the mean $E(b-y)$ is found to be 0.070 $\pm$ 0.003 (sem). The second intrinsic color relation is that of
\citet{nis88}, a slightly modified version of the original relations derived by \citet{craw75, craw79} for F and A stars. For the
same 61 F dwarfs, the alternate relation implies $E(b-y)$ = 0.075 $\pm$ 0.003 (sem). Because of its age, 
NGC 5822 has a rich population of A stars. Using the \citet{nis88} relation for the stars with H$\beta$ above 2.72, from 
48 stars one finds $E(b-y)$ = 0.078 $\pm$ 0.002 (sem). It should be noted that the slightly higher reddening for F stars
using the \citet{nis88} relation compared to that of \citet{ols88} is a consistent occurrence from such 
comparisons \citep{tcat06, attm07}. A weighted average of all three results leads to $E(b-y)$ = 0.075 $\pm$ 0.003, or 
$E(B-V)$ = 0.103 $\pm$ 0.003 (sem), which we will adopt for the cluster in all future discussions. When combined with the
zero-point uncertainties in $b-y$ and H$\beta$, the total uncertainties in $E(b-y)$ and $E(B-V)$ become $\pm$ 0.008 mag
and 0.011 mag, respectively. 

With the reddening fixed, the next step is the derivation of metallicity, a parameter that can be defined using $hk$ or
$m_1$ coupled to either $b-y$ or H$\beta$ as the primary temperature indicator. In past studies using $uvbyCa$H$\beta$ photometry,
the metallicity from $hk$ tied to H$\beta$ invariably has been given the greatest weight due to the greater sensitivity of $hk$ to
modest metallicity changes, while the H$\beta$-based relations allow decoupling between errors in the two indices and minimize the impact
of potential reddening variations, if any exist. We will follow the same approach with NGC 5822, allowing us to tie our results directly into the
same metallicity scale generated in past intermediate-band cluster studies.

With $E(b-y)$ = 0.075, the mean $\delta$$m_1$($\beta$) for 61 F dwarf probable members between H$\beta$ = 2.58 and 2.72 is 0.017 $\pm$ 0.003 (sem),
where $\delta$$m_1$ = 0.0 is set at the adopted Hyades metallicity of +0.12. On this same scale, NGC 3680 and IC 4651 have $\delta$$m_1$ =
+0.027 $\pm$ 0.002 (sem) \citep{att04} and 0.000 $\pm$ 0.002 (sem) \citep{att00}, implying that NGC 5822 is clearly lower in [Fe/H] than the Hyades, 
but not as deficient as NGC 3680. The comparison can be improved because the slope of the $\delta$$m_1$ - [Fe/H] relation is color-dependent 
\citep{nis88} and the sample of F dwarfs in NGC 5822 is more heavily weighted toward hotter stars than in NGC 3680. For NGC 5822, the 
$\delta$$m_1$ measure translates to [Fe/H] = -0.058 $\pm$ 0.027 (sem), on a scale where NGC 3680 and IC 4651 have [Fe/H] = -0.175 and +0.115, 
respectively. As an additional reference point, the photoelectric $uvby$H$\beta$ data of M67 produce [Fe/H] = -0.06 \citep{ntc87}. 

Turning to the $hk$ index, $\delta$$hk$($\beta$) = 0.031 $\pm$ 0.006 (sem), which translates to [Fe/H] = 0.010 $\pm$ 0.020 (sem), on a scale
where [Fe/H] = +0.12 for the Hyades and NGC 3680 has [Fe/H]$_{hk}$ = -0.105 $\pm$ 0.016 (sem). A weighted average of the two metallicity estimates
leads to [Fe/H] = -0.019 $\pm$ 0.023, where the errors refer to the internal errors from the photometry alone.

As a modest consistency check on our scale, we turn to one of the more recent attempts to redefine the metallicity calibration for the
$uvby$ system for dwarfs over a range in temperature from F through K stars. \citet{hol07}(HNA) discuss the issues with the
calibration of \citet{sn89}, which served as a starting point for the calibrations adopted by \citet{nor04}. The net result of the
2004 analysis was a series of metallicity relations covering different color ranges, including a $(b-y)$-based, F-dwarf relation for stars 
between $b-y$ = 0.18 and 0.38, that make direct use of the indices without reference to a standard relation, following the lead of
\citet{sn89}. Equally important for the current discussion is the lack of a $c_1$ dependence among the terms used in the calibration, a point we
will return to below. If we correct the NGC 5822 stars for reddening of $E(b-y)$ = 0.075 and apply the relation of \citet{nor04} to 66 
F dwarfs, the mean [Fe/H] = -0.054 $\pm$ 0.031 (sem) on a scale where the Hyades produces [Fe/H] = +0.10. 

Since the fainter photometry for the unevolved main sequence of NGC 5822 extends to $(b-y)­_0$ redder than 0.38, it could be useful to derive
[Fe/H] for the G-dwarfs within the cluster. The problem with this approach is the increasing uncertainty in the $c_1$ indices at fainter magnitudes
and redder colors due to the absence of cooler dwarfs within the calibration from the instrumental to the standard system. Comparison of the dereddened
indices to the fiducial $c_1$ -$(b-y)$ relation shows that the difference between the two, in the sense (OBS - FID), starts off positive at the blue end,
as expected since the stars in the turnoff region are evolved, then declines to zero near $(b-y)­_0$ = 0.33. Redder than this color, the offset should
be 0.0 or slightly negative with increasing color. The expected negative $\delta$$c_1$ among the late F and early G stars is a reflection of the increased 
metallicity sensitivity of $c_1$ and the declining sensitivity of $m_1$ at a given $b-y$ \citep{ols84, ntc87, tatt02, tvat07}, coupled to the less than Hyades
metallicity of NGC 5822. However, at redder $b-y$ the cluster $c_1$ photometry continues to decline at a faster rate than the fiducial relation, enhancing
the negative $\delta$$c_1$ values. For the metallicity determinations discussed above, this trend has no impact because the relations are defined 
for the hotter stars and have no $c_1$ dependence. By contrast, for stars redder than $(b-y)_0$ = 0.30, if one adopts an indices-defined 
relation as found in HNA, where 10 of 20 terms include $c_1$ in some form, or in \citet{cas11}, where 4 of 11 terms include $c_1$ in 
some form, an error in $c_1$ can have a significant impact on [Fe/H]. The pattern is illustrated in Fig. 15 where the newly revised $uvby$ 
metallicity calibrations for intermediate ($(b-y)_0$ $< 0.43$) and redder colors ($(b-y)_0$ $>$ 0.43) have 
been applied to 55 probable members of NGC 5822 (open circles). The mean metallicity from 13 stars with $(b-y)_0$ $<$ 0.34 is -0.21 $\pm$ 0.11 (sd), 
in contrast with [Fe/H] = -0.06 from the F-star relation used above and, as one moves redward, the predicted [Fe/H] declines in a linear fashion, 
reaching a mean near -0.75 at $(b-y)_0$ of 0.50. A virtually identical pattern is produced using the relations in HNA.

As discussed in \citet{tvat07}, one can reduce the impact of random and systematic photometric errors in $c_1$ on [Fe/H] if H$\beta$ 
photometry is available for the cooler dwarfs. The
crosses in Fig. 15 show the results for the same 55 dwarfs if the H$\beta$-based metallicity relations derived in \citet{tvat07} are applied. 
The abundance scale adopted in \citet{tvat07} is that of \citet{vf05}. Based upon the revision of the HNA $uvby$ metallicity calibration by \citet{cas11},
which raised the mean [Fe/H] of the HNA scale by $\sim$ 0.1 dex, the two systems in Fig. 15 should be similar, if not identical. 
However, the H$\beta$-defined abundances for the same 13 stars at the blue end of the sample generate [Fe/H] = -0.10 $\pm$ 0.09 (sd) and, while the 
general trend of declining [Fe/H] with increasing $(b-y)_0$ still applies, the slope is shallower and the scatter at a given $(b-y)_0$ is significantly
smaller. From 24 stars redder than $(b-y)_0$ = 0.40, the \citet{cas11} calibration systematically 
underestimates [Fe/H] by 0.23 $\pm$ 0.11 (sd) dex relative to the H$\beta$-defined relation.  

\subsection{Comparison to Previous Results}
The reddening and metallicity estimates for NGC 5822 as of 1993 are discussed in detail in \citet{tam93}. More recent results are
presented in Table 1. For reddening, DDO photometry of 16 member giants produced $E(B-V)$ = 0.143 $\pm$ 0.012 (sem), while the $uvby$H$\beta$ 
analysis of 21 turnoff stars \citep{stet87} indicated $E(B-V)$ = 0.139 $\pm$ 0.009 (sem). Other analyses produced reddening values ranging 
between 0.11 to 0.19, but these invariably contained non-members and/or were subject to large uncertainties. The only redetermination of 
the reddening since then has been that of \citet{pac10}, obtained
by matching the CMD defined by photoelectric measurements to a set of isochrones with the metallicity derived from spectroscopic analysis of 2 dwarfs.
The uncertainty is large, but the result, $E(B-V)$ = 0.1 $\pm$ 0.05, is consistent with the value derived in this investigation.

For metallicity, the dominant contributors to the adopted cluster average in \citet{tam93} were DDO photometry ([Fe/H] = -0.11)
and $UBV$ photometry of the giants and dwarfs ([Fe/H] = -0.15). We will rediscuss the $UBV$-based reddening and metallicity in Sec. 5.
The only spectroscopic results available were the moderate-dispersion 
spectroscopic analyses of \citet{fj93} for 3 giants, including 2 binaries, leading to [Fe/H] = -0.21. With the rederivation of the DDO
metallicity scale and the rescaling of the moderate-dispersion spectroscopic data of \citet{fj93}, \citet{tat97} found [Fe/H] = -0.03 $\pm$ 0.02 (sem)
from 17 giants. Since the reddening has been lowered to $E(B-V)$ = 0.10, some adjustment of this value is required. Fortunately, the impact 
of lowering the reddening works in opposite directions by raising [Fe/H] from DDO but lowering the spectroscopic estimates. The adjusted 
value becomes [Fe/H] = -0.01 on a scale where NGC 3680, M67, and IC 4651 are -0.10, 0.00, and +0.10, respectively.

Until recently, the only high-dispersion spectroscopy of NGC 5822 included three giants, one of which is a definite non-member \citep{luc94}. The
average [Fe/H] for the two members, including one binary and a potential AGB star, is +0.06 $\pm$ 0.03, but the adopted temperature may be $\sim$ 300 K
too hot for one of the stars \citep{smi09}. If the temperature scale is based purely upon the photometric colors, the mean [Fe/H] is lowered to -0.06. 
From 5 giants, \citet{smi09} find [Fe/H] = +0.04 $\pm$ 0.08 (sd) using spectroscopic temperatures that are, on average, only 37 K cooler 
than the photometric temperatures derived under the assumption that $E(B-V)$ = 0.14. With the adopted lower reddening, the photometric temperatures
would be 80 K lower \citep{hou00}; an additional drop of 40 K in the adopted temperature scale would lower the spectroscopic abundance by 0.04 dex. 
\citet{sa09} analyze 3 giants, none of which overlap with the earlier work, and find [Fe/H] = 0.05 $\pm$ 0.04 or [Fe/H] = +0.12 $\pm$ 0.10, 
depending upon the adopted line list. Finally, the only dwarfs studied to date are two stars noted earlier in the sample of \citet{pac10} which 
generate [Fe/H] = +0.05 $\pm$ 0.03; a rediscussion of the same three giants in \citet{sa09} gives [Fe/H] = 0.15 $\pm$ 0.08.

\section{Cluster Properties - Age and Distance}
With the reddening and metallicity known, we now turn to the determination of the cluster age and distance through comparison to theoretical isochrones.
To optimize the fits, we would prefer to use only single-star members which, with the eventual exception of some key stars among the red giants, 
restricts the sample to stars with both broad-band and intermediate-band photometry, i.e. the central 13.5\arcmin$\times$25\arcmin\ of the cluster. 
As detailed in Sec. 3, we can remove virtually all background stars due to the dramatic increase in reddening beyond the cluster. We can also 
isolate highly probable foreground stars and
cluster binaries with a mass ratio near 1.0 for $V$ fainter than 12.5 by identifying stars that deviate excessively from the mean relations in the 
$V, b-y$ and $V, hk$ CMD's, as illustrated in Figs. 12 and 13. Both approaches become less reliable among hotter stars near the vertical turnoff because
binarity among stars in this region shifts the composite system into an area also occupied by evolved single stars. One additional CMD remains which
may allow additional composite interlopers to be identified, the ($V, c_1$) diagram. For unevolved stars, $c_1$ increases steadily as one moves up the main
sequence, with a range of over 0.65 mag from unevolved early G stars to early A stars. Additionally, evolution away from the main sequence 
causes a correlated increase in $c_1$ at a given $b-y$ as $V$ becomes brighter. Thus, stars at the vertical turnoff should follow a well-defined 
trend of larger $c_1$ at brighter $V$. Binaries should reveal themselves by being too bright at a given $c_1$. 

Fig. 16 shows the ($V, c_1$) plot for the same stars found in Figs. 11 and 12, with symbols having the same meaning. Six additional stars (star symbols) 
brighter than $V$ = 12.5 have been identified as probable binaries because their positions in the CMD are consistent with binarity but 
incompatible with single stars affected by plausible photometric errors in either $V$ or $c_1$. Note that the
deviant points in this figure at fainter magnitudes have already been identified as such in one of the previous comparisons. In all comparisons of
the cluster to theoretical isochrones, any star tagged as deviant as illustrated in Fig. 16 will be excluded from the discussion.

For the initial comparison, we use the broad-band ($V, V-I$) CMD, primarily because it has the smallest photometric errors to the faintest magnitude among
the $UBVI$ indices. The single stars of Fig. 16 have been matched to the data of Table 3, generating a sample of 145 stars bluer than $V-I$ = 0.7. 
For the evolved region of the CMD, a list of all stars with $V-I > 0.7$ and $V$ brighter than 12.0 was compiled from Table 3. This list was 
matched with the radial-velocity results of \citet{mer89} and \citet{mm90}, as summarized in \citet{mer08}. Of the 28 stars included in the radial-velocity 
study, 20 are found in Table 3; the remaining 8 include four nonmembers and four members brighter than $V$ = 9.75. The 20 stars for which $VI$ 
data are available include 8 single-star members, 8 binary members and 4 nonmembers. The remaining
stars redder than 0.7 were matched with the intermediate-band data of Table 5. Three stars were easily identifiable as heavily reddened background
stars, 4 had indices consistent with low reddening, implying cluster members or foreground stars, and 13 were located outside the area covered by
the intermediate-band survey. Excluding non-members, the entire sample is plotted in Fig. 17. For $V-I$ below 0.7, open circles represent probable
single-star members based upon intermediate-band photometry. For $V-I > 0.7$, open circles are single-star, radial-velocity members, open triangles
are radial-velocity member binaries, starred points are stars with low reddening implied by intermediate-band photometry, and crosses are stars with only
$VI$ data. 

Superposed are the isochrones with [Fe/H] = 0.00 and ages of 0.8, 0.9 and 1.0 Gyr, shifted by E$(V-I)= 1.35\ {\rm E}(B-V)$ = 0.139 and an apparent modulus of 
$(m-M)$ = 9.85. For consistency with the comparable discussions of NGC 3680, NGC 752, and IC 4651 in \citet{att09}, the isochrones and interpolation 
software for specific ages and abundances adopted for the present discussion are those of Y$^{2}$ (http://www.astro.yale.edu/demarque/yyiso.html) 
\citep{yi03,dem04}. A modest adjustment of -0.03 mag has been applied to the isochrone $M_V$ scale to make the solar models compatible with 
our consistently adopted value of $M_V$ = 4.84. It should be noted that a shift of this size is larger than the range expected due to the uncertainty 
in the determination of [Fe/H] \citep{tae09}. No adjustment has been made to the $V-I$ scale. Giving heavy weight to the color of the turnoff, 
obvious blue stragglers excluded and the uncertainty in the reddening included, the cluster age is well constrained at 0.90 $\pm$ 0.10 Gyr. 

Moving away from the turnoff region toward the giants, the group of seven stars populating the Hertzsprung gap is likely dominated by foreground field stars rather than stars transitioning to the giant branch given that (a) they lie well below the subgiant branch, (b) this phase
is rapid enough that no more than one star should be captured during this transition and (c) most of these stars lie outside the cluster
core region, as expected for a field star distribution. By contrast, despite the lack of four known members in the photometric sample, the giant
branch is richly populated, with five new potential members stars falling among the regions dominated by the radial-velocity members. Four of the
five new candidates are equally split between the already discussed dual $clumps$ at $V$ = 10.35 and 10.8. With the expanded
sample and the difference in the color spread among the two clumps, the red giant branch now takes on a more traditional appearance, with the brighter and
broader clump at $V$ = 10.25 to 10.45 and $V-I$ = 1.02 to 1.11 potentially associated with core He-burning stars and the fainter, tighter group 
($V-I$ = 1.05 to 1.09) populated by first-ascent red giants. In the absence of radial-velocity information clarifying membership and binarity, 
this interpretation should be regarded with caution. We will return to this point in discussing the second CMD comparison, the ($V, B-V$) diagram. 

For the $BV$ CMD analysis, a much larger database exists beyond the photometry of Table 3. As discussed in Sec. 2, NGC 5822 was observed photoelectrically
on a number of occasions prior to the CCD era, with a predominant emphasis on the brightest stars in the field for all studies except \citet{tam93}. As the
comparisons of Sec. 2 also demonstrate, the majority of these studies compare favorably with the CCD data in terms of photometric precision. To make optimal
use of this data, the photometry of \citet{bsm68,jh75,cl85,cl89,mer08} was adjusted to the photoelectric system of 
\citet{tam93}. The photometry of Table 3 was adjusted by +0.003, +0.007, and -0.010 in $V$, $B-V$, and $U-B$, respectively, to place it on the
photoelectric system. Note that the offsets are so small that adoption of either system for the zero-points has a negligible impact on our
conclusions. Finally, the high internal precision of the $by$ photometry allows us to readily transfer this data to the $BV$ system with little need
to be concerned about distorting the CMD relations. Using only stars classed as single-star cluster members along the main sequence, i.e. excluding
stars classed as blue stragglers, subgiants or giants, the $(B-V)$ versus $(b-y)$ data were fit with two linear relations:

\noindent
$(b-y)$ $\leq$ 0.381 \ \  $ B-V = (0.147 \pm 0.011) + (1.015 \pm 0.038) \times (b-y)$ \\
$(b-y)$ $>$ 0.381 \ \  $ B-V = (-0.227 \pm 0.015) + (1.996 \pm 0.030) \times (b-y)$ \\
\noindent
The dispersions among the residuals for the transformed photometry are 0.039 and 0.034 mag, respectively. The change in slope near $B-V$ = 0.53 is
sharp and significant, a point we will return to below.

The individual data sets were assigned a weight based upon their comparison to the adopted photoelectric set and averaged. The stars in Fig. 17
were selected and plotted in Fig. 18 using the same symbols, reddening, and distance modulus as in Fig. 17, adopting the same set of 
isochrones used, but on the $BV$ system. As with Fig. 17, the fit of the isochrones is quite good, though not perfect. The lower main sequence does an excellent job
of matching the observations, but the isochrones deviate from the unevolved main sequence at a steeper rate than the cluster stars. At the turnoff,
the age of the cluster defined in the region of the red hook is slightly older than that from the $VI$ CMD, but the same within the uncertainties in the
colors and the profile of the isochrones at 0.9 $\pm$ 0.1 Gyr. As before, all the stars on the subgiant branch at intermediate colors lie well below
the predicted location of the isochrone and should be regarded as probable non-members. The exception is the known binary member, which has a luminosity
and color indicative of a turnoff-red-giant pair.

Moving to the red giant branch, the two faintest stars once again lie redward of the predicted isochrone and, while it is tempting to dismiss these
stars as non-members, sub-subgiants are known to exist in older open clusters, M67 providing the best examples \citep{mat03}. If they are members, however,
these stars are anomalies, potentially binaries, by definition and tell us nothing about normal red giant evolution. Moving along the more 
traditional red giant track, the giant branch once again breaks into two distinct regions, with a tight grouping of stars at fainter magnitudes 
and a broader clump of stars approximately 0.5 mag brighter. The fainter group superposes nicely on the first-ascent red giant branch, while 
the brighter group scatters systematically to the
red. The one probable new member of the red giant clump with both intermediate and broad-band observations falls significantly redward of the giant 
branch in Fig. 18, but its positions in Fig. 17 and  Fig. 9 place this star at the level of the clump on or blueward of the first-ascent giant branch, 
making it likely that the extreme position in Fig. 18 is a product of a modest photometric error. 

Are the two groupings in the giant branch simply a separation of stars on the first-ascent and second-ascent giant branches? While 
post-He-core-ignition tracks are unavailable for the isochrones used in Figs. 17 and 18, we can turn to the stellar models of \citet{gir01}
to evaluate the expected distribution of He-core-burning giants, following the example of \citet{gir00b} using the
older models of \citet{gir00a}. Fig. 19 shows the predicted limiting (faintest) locations for core-He-burning
stars as a function of their initial mass (solid line), adjusted for the reddening and distance of NGC 5822; symbols have the same
meaning as in Fig. 17. Evolved stars of $\sim$2.0 solar masses superpose nicely on the fainter clump of stars, but the brighter clump seems to
require a spread in mass reaching to $\sim$2.5 solar masses. A distribution of 0.6 mag or more in $M_V$ over a narrow range in color among the 
clump stars is consistent with the synthetic CMDs detailed in \citet{gir00b} for ages from 0.8 to 1.0 Gyr, but no break occurs in the predicted 
distributions. Assuming the break between the clumps is not simply a statistical fluctuation, the failure of the models to create two distinct 
clumps, as also claimed for NGC 752 and NGC 7789, has led in part to the suggestion of an extended star formation history within the individual
clusters or potentially an indication of the transition to degenerate-He-core ignition among the lowest mass stars on the giant branch. The 
latter explanation seems unlikely for a cluster as young as NGC 5822. NGC 5822 therefore provides a key test case at an age where the 
two observed clumps are equally populated, unlike NGC 7789 and NGC 752 at ages between
1.4 Gyr and 1.8 Gyr, where the fainter clumps are, at best, weakly populated and a challenge to distinguish from first-ascent red giants which 
should be observable below the clump.

\section{Reddening and Metallicity Revisited: UBV Photometry}
To close the discussion of the cluster properties, we revisit the broad-band $UBV$ data for the unevolved main sequence stars of Fig. 18. 
All stars fainter than $V$ = 12.0 were sorted in $B-V$ in bins 0.05 mag wide starting at $B-V$ = 0.35. The mean values of $U-B$ versus $B-V$
are plotted in Fig. 20; error bars represent the dispersion in $B-V$ and the standard error of the mean in $U-B$. Superposed is the standard Hyades
two-color relation \citep{san69} shifted, from left to right in the plot by $E(B-V)$ = 0.05, 0.10, and 0.15. $E(U-B)$ has been 
derived assuming $E(U-B)$/$E(B-V)$ = 0.72 + 0.05$E(B-V)$.

The first feature of the figure requiring comment is the steep slope of the cluster data in contrast with the standard relations for $B-V < 0.55$.
The more rapid growth of $U-B$ with declining $B-V$ forces the cluster relation to cross the Hyades relation for $E(B-V)$ below 0.10 and, if the
standard relations were extended to bluer colors, probably for even higher reddening values. Since the points between $B-V$ = 0.6 and 0.8 all sit
above the Hyades relation, independent of the adopted reddening, this transition to super-Hyades metallicity seems unlikely. A more plausible solution
is tied to the rather sharp change in the slope of the $(b-y)$ - $(B-V)$ relation near $B-V$ = 0.53 noted earlier. It has been known for decades
that, in addition to the size of the color shift in $U-B$ above the standard relation being a function of $B-V$ for a given change in [Fe/H], the
shift is affected by evolution off the main sequence for stars bluer than $B-V$ = 0.55 \citep{es64}. The greater the degree of evolution, the redder
$U-B$ appears, with the size of the effect dependent upon $B-V$; the increase in $U-B$ for a given change in magnitude grows as $B-V$
shifts from 0.55 to 0.45 and remains relatively constant to $B-V$ $\sim$ 0.25 before declining again. Therefore, we conclude that the steep slope in the
two-color diagram and the change in slope for the $(b-y)$ - $(B-V)$ relation beginning at $B-V$ = 0.53 ($(B-V)_0 = 0.43$) reflects the increasing 
degree of evolution of stars off the main sequence as $B-V$ declines. Stars bluer than $B-V$ = 0.55 should not be used to derive the reddening and/or
metallicity. This explains why the analysis of \citet{tam93} generated a reddening value ($E(B-V)$ = 0.15) which now appears to be too large. The
$UBV$ photoelectric photometry is heavily weighted by stars bluer than $B-V$ = 0.50 within the turnoff region affected by evolution. To obtain any
estimate of an ultraviolet excess that placed the cluster below the metallicity of the Hyades, as required by the few redder dwarfs with
photoelectric data, demanded $E(B-V)$ much greater than 0.10 to compensate for the steepened two-color trend for the bluer stars. 

At the red end of the scale where evolution effects should be absent, we can place another constraint on the reddening. For $E(B-V)$ = 0.15 the
cluster data crosses the two-color relation for $B-V$ = 0.9; all points redder than this boundary sit below the Hyades relation, again in contradiction
with the points at intermediate color. To have the redder cluster data lie at a lower metallicity than the Hyades requires than $E(B-V)$ be no greater
than 0.12.

Given the three options for reddening in Fig. 20, what metallicity is implied by the two-color data? Using only the mean data redder than $B-V$ =
0.60, the ultraviolet excesses were calculated for each point and corrected for the color dependence of the ultraviolet excess to transform the mean
value to the determination if all data had $(B-V)_0$ = 0.6, i.e. $\delta$$_{0.6}$. The simple averages for $E(B-V)$ = 0.05, 0.10, 0.15 
are 0.113 $\pm$ 0.044 (s.d.), 0.062 $\pm$ 0.031 (s.d.), and 0.018 $\pm$ 0.046 (s.d.), respectively. Note that the growth in the dispersion
for values on either side of $E(B-V)$ = 0.10 is a reflection of the larger range in $\delta$$(U-B)$ created by a color shift in the standard relation
which has a more significant impact on the redder dwarfs than stars of intermediate color. The minimum in the dispersion for $E(B-V)$ = 0.10 is
an indicator that the collective trend among the cluster data optimally matches the profile of the standard relation.

A variety of relations have been derived over the years to transform from $\delta$$_{0.6}$ to [Fe/H] \citep{wal62, car79, cam85, sf87}. For 
convenience we make use of one of the more recent discussions \citep{kar06}, adopting their
final relation tied to 266 dwarfs and turnoff stars, which should have the greatest applicability to the data in Fig. 20. On 
a scale where the Hyades, with $\delta$$_{0.6}$ = 0.0, would have [Fe/H] = +0.09, the three metallicity estimates for NGC 5822 
become [Fe/H] = -0.46, -0.16, and +0.03. If we had adopted the $U-B$ scale of the CCD survey, shifting $U-B$ values by +0.010 mag raises 
the three values to -0.40, -0.11, and +0.07, but also places an even tighter constraint on the upper limit for the allowed reddening if
we demand that the mean data for the cluster not cross the Hyades relation among the redder dwarfs. We conclude that the expanded and
improved $UBV$ data for NGC 5822 require a slightly lower reddening for the cluster than the original analysis by \citet{tam93}, with a
most probable range between $E(B-V)$ = 0.10 and 0.125 and a coupled [Fe/H] between -0.16 and 0.00, taking into account the uncertainties
in the $U-B$ zero point and the zero-point of the metallicity calibration.

\section{Summary}
The intermediate-age open cluster, NGC 5822, has been re-evaluated using an extensive broad-band survey covering an area $\sim$40\arcmin\
on a side, complemented by a precision $uvbyCa$H$\beta$ study of the core area of the cluster. The latter sample clearly shows that the
nearby cluster is superposed upon a background field of significantly higher reddening than the cluster, making identification and isolation
of probable cluster members a straightforward exercise. After photometric elimination of probable foreground and/or binary stars, a
consistent reddening of $E(b-y)$ = 0.075 $\pm$ 0.008 mag or E$(B-V)$ = 0.103 $\pm$ 0.011 mag is derived from analysis of 48 A and 61 F dwarfs.
With the reddening defined, metallicity estimates from the $m_1$ and $hk$ indices imply an effectively solar to slightly subsolar metallicity, consistent
with the metallicity from DDO photometry of the rich giant population and some high-dispersion spectroscopic analyses, though not all. Despite a
lower reddening and slightly higher metallicity than found in the previous CMD analysis using photographic photometry, the cluster retains the
same apparent modulus, $(m-M)$ = 9.85, but a younger age (0.9 $\pm$ 0.1 Gyr) than derived by \citet{tam93}, a byproduct of using 
significantly improved isochrones.

How do the lower age and reddening impact previous analyses based upon inclusion of cluster members with the previous parameters? Two
affected studies should be noted. With the removal of probable field stars and binaries from the CMD and a lower reddening, the 
B{\"o}hm-Vitense gap at $B-V$ = 0.52 proposed by
\citet{rc00} now lies directly over the predicted location for the expected Li-dip in NGC 5822 at $(B-V)_0$ = 0.41 \citep{att09}, though
the significance of the gap now seems less apparent. It should be noted that the one apparent break in the distribution of stars with $V$ rather than
$B-V$ that remains from the photographic study is the decline in probable members between $V$ = 11.5 and 12.0. The $(B-V)_0$ color of this
break for a less evolved cluster would place the decline in the 0.20 to 0.25 range, more consistent with the original color location of the
gap as defined by \citet{bc74}.  

A more direct impact of the younger age is a partial alleviation of the need for a
dramatic transition from chromospherically active to inactive solar-type stars near the age of 1.3 $\pm$ 0.1 Gyr. \citet{pac09}, using observations
of 2 solar type stars in NGC 5822, found the stars to be as active as stars of similar mass within the Hyades, Praesepe, and IC 4756, all clusters with 
ages below 1 Gyr. By contrast, solar-type stars in NGC 3680 and IC 4651 both exhibited low activity levels. With an adopted age of 1.4 Gyr for 
NGC 3680 and 1.2 Gyr for NGC 5822, they concluded that the physical process driving the activity underwent a dramatic decline over
a period of $\sim$0.2 Gyr or less. Using the same isochrone set and photometric approach adopted for NGC 5822, \citet{att09} derived ages of 
1.5 Gyr and 1.75 Gyr for IC 4651 and NGC 3680, implying a gap of $\sim$0.6 Gyr between the chromospherically active and inactive stars.

Finally, the expanded photometric sample has added a number of potential members to the giant branch, including four stars split evenly between
the two distinct clumps that define the red giant distribution in open clusters in this age range. The balanced population of the two clumps
is unique among open clusters of this age range and, coupled with the distinct break in $V$ between the two groups, is difficult to explain under 
any scenario involving normal, single-star evolution of a given age. To decide if the bimodality is a product of an extended red giant clump driven
by a range in mass among the stars leaving the main sequence or a failure of standard stellar evolution to predict the correct distribution
of stars on the first-ascent versus second-ascent red giant branch requires a means of deciding which phase the two groups represent,
beyond the simple criterion of location within the CMD. Given that the stars at the turnoff all lie well blueward of the Li-dip, the one
obvious means of potentially distinguishing between the two scenarios is a measure of the Li abundances of the two groups; second-ascent
(He-core-burning) red giants should be seriously depleted in Li relative to the first-ascent members, which should exhibit a decline relative 
to the stars at the turnoff.

\acknowledgments
The authors express their sincere thanks for the thoughtful comments of the referee which helped clarify the issues raised in the analysis.
Extensive use was made of the WEBDA database maintained by E. Paunzen at the University of Vienna, Austria (http://www.univie.ac.at/webda). 
The filters used in the program were obtained by BJAT and BAT through NSF grant AST-0321247 to the University of Kansas. BAT gratefully 
acknowledges travel support to CTIO provided by the University of Kansas Travel Fund. EC acknowledges support by the Fondo Nacional de Investigaci\'on
Cient\'{\i}fica y Tecnol\'ogica (proyecto No. 1110100), the Chilean Centro de Astrof\'{\i}sica (FONDAP No. 15010003) and the Chilean Centro de 
Excelencia en Astrof\'{\i}sica y Tecnolog\'{\i}as Afines (PFB 06). BAT is grateful to the administration at ESO for 
their hospitality during an initial visit at the start of this project, while GC acknowledges ESO support for an extended visit to the 
University of Kansas. BJJ gratefully acknowledges the support of the NSF through grant AST 08-0850564 as part of the CSUURE Research 
Experiences for Undergraduates (REU) program at San Diego State University.

\begin{figure}
\epsscale{.80}
\centering
\plotone{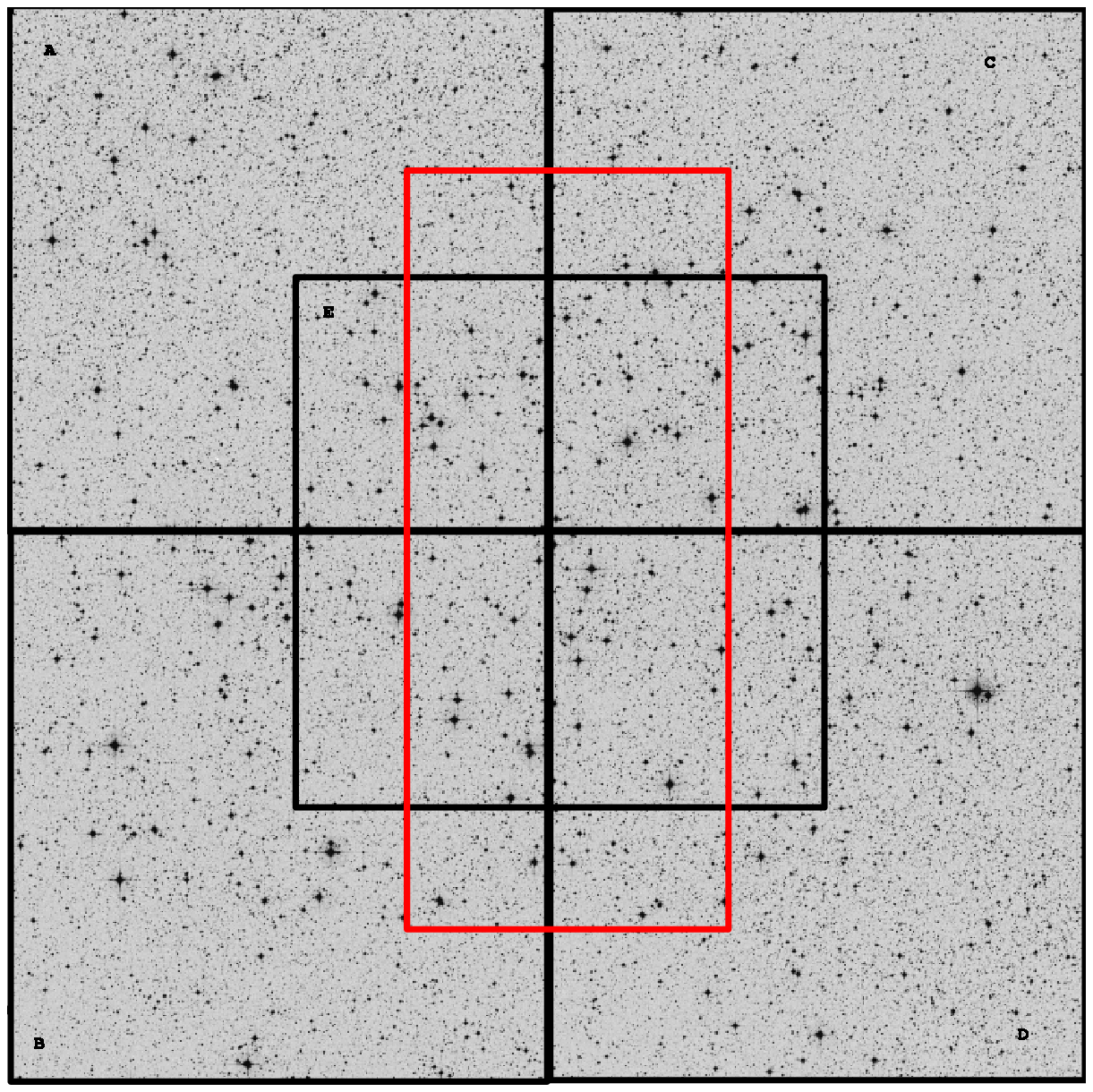}
\caption{DSS image centered on NGC 5822, illustrating the five Y4KCam pointings. North is up, East to the left, 
and the field of view is 40 arcmin on a side.}
\end{figure}

\clearpage
\begin{figure}
\includegraphics[width=\columnwidth,angle=270,scale=0.80]{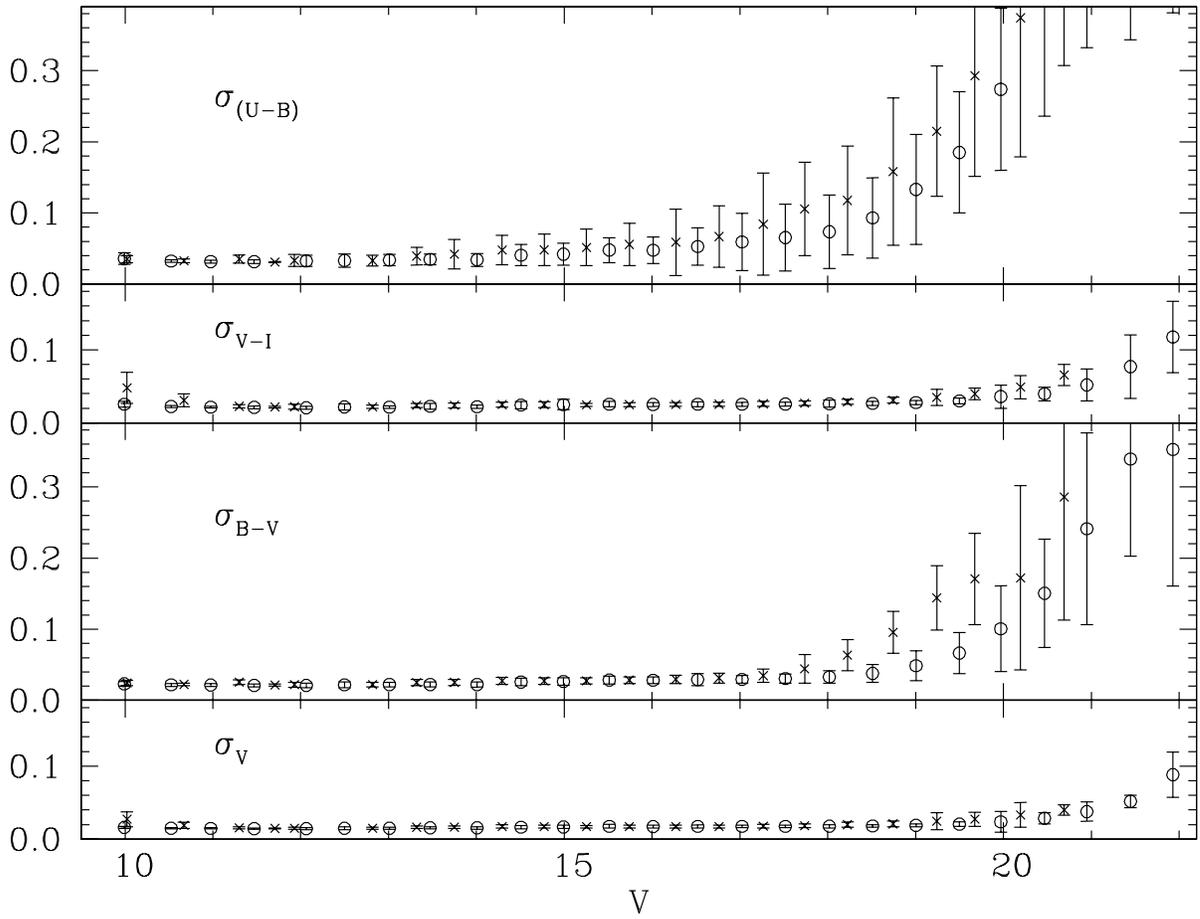}
\caption{Average photometric errors in $V$, $(B-V)$, $(U-B)$, and $(V-I)$ as a function of the $V$ magnitude. Open
circles refer to stars within region E, while crosses represent data for stars outside of E.}
\end{figure}

\clearpage
\begin{figure}
\includegraphics[width=\columnwidth,angle=270,scale=0.80]{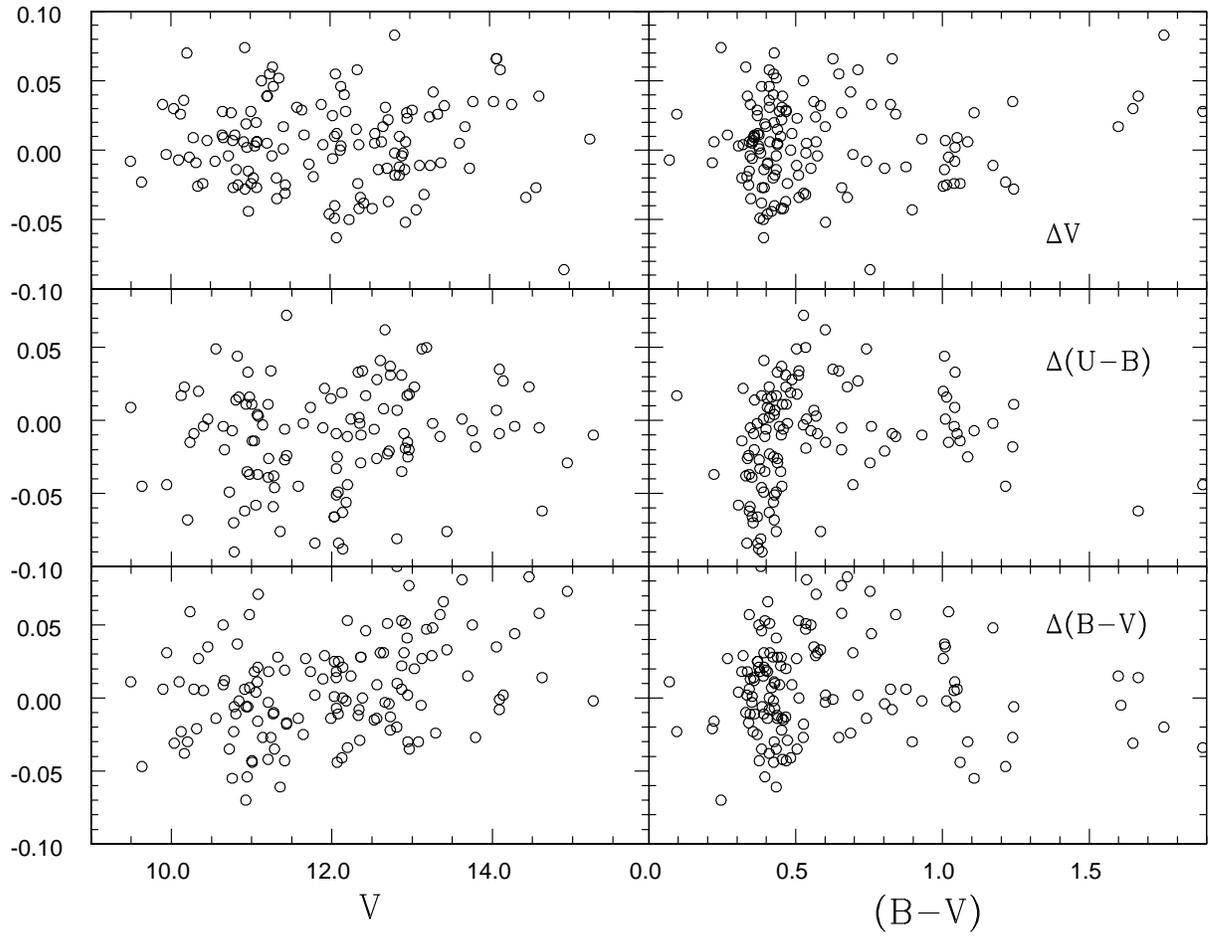}
\caption{Residuals, in the sense (TAM - Table 3), between the photoelectric and CCD observations for $V$, $B-V$, and
$U-B$ as a function of $V$ and $B-V$.}
\end{figure}

\clearpage
\begin{figure}
\includegraphics[scale=0.80]{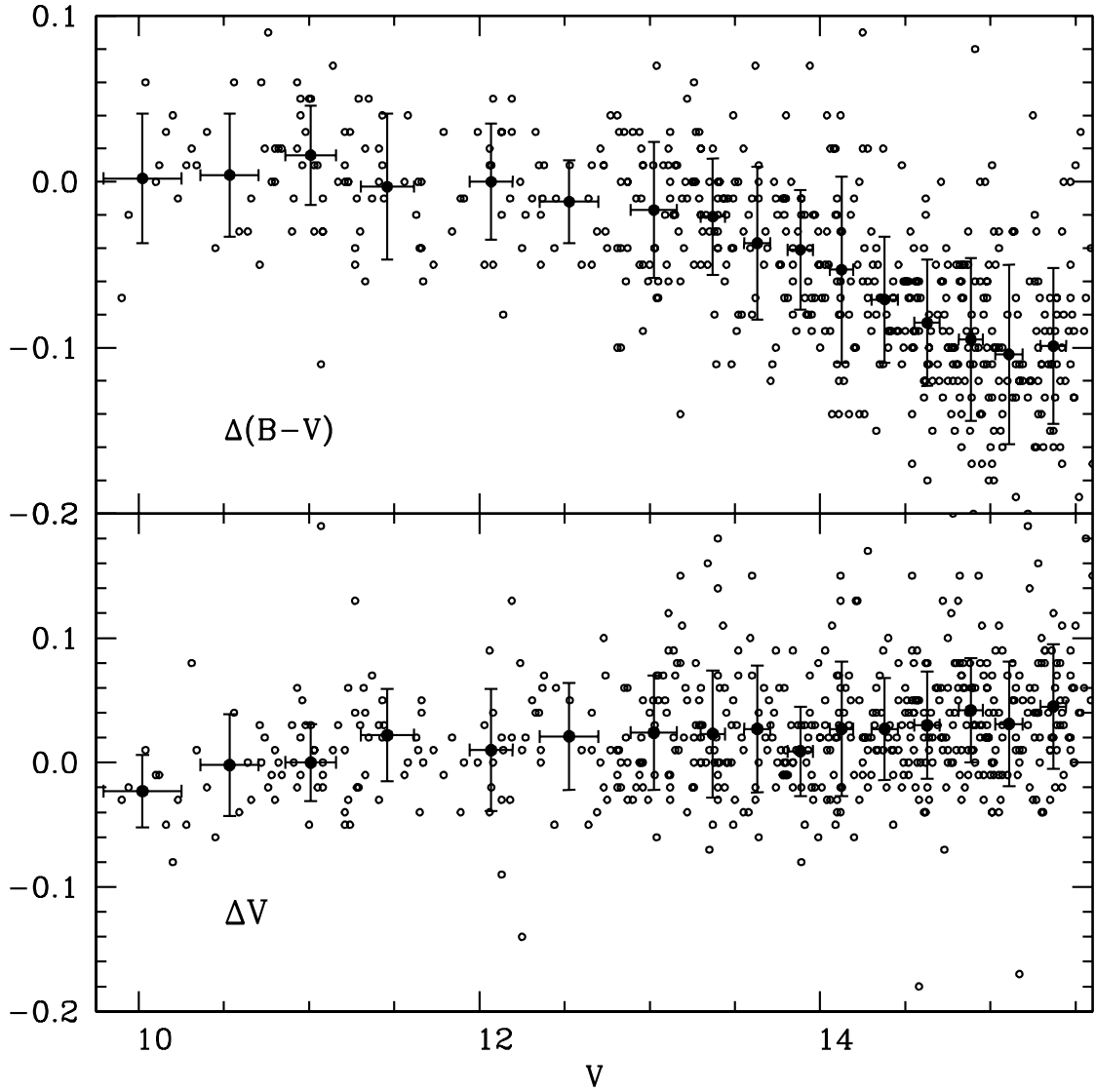}
\caption{Residuals, in the sense (Table 3 - PG), between the photographic and CCD observations for $V$ and $B-V$ as 
a function of $V$. Filled circles represent the average values within a magnitude bin.}
\end{figure}

\clearpage
\begin{figure}
\includegraphics[width=\columnwidth,angle=270,scale=0.80]{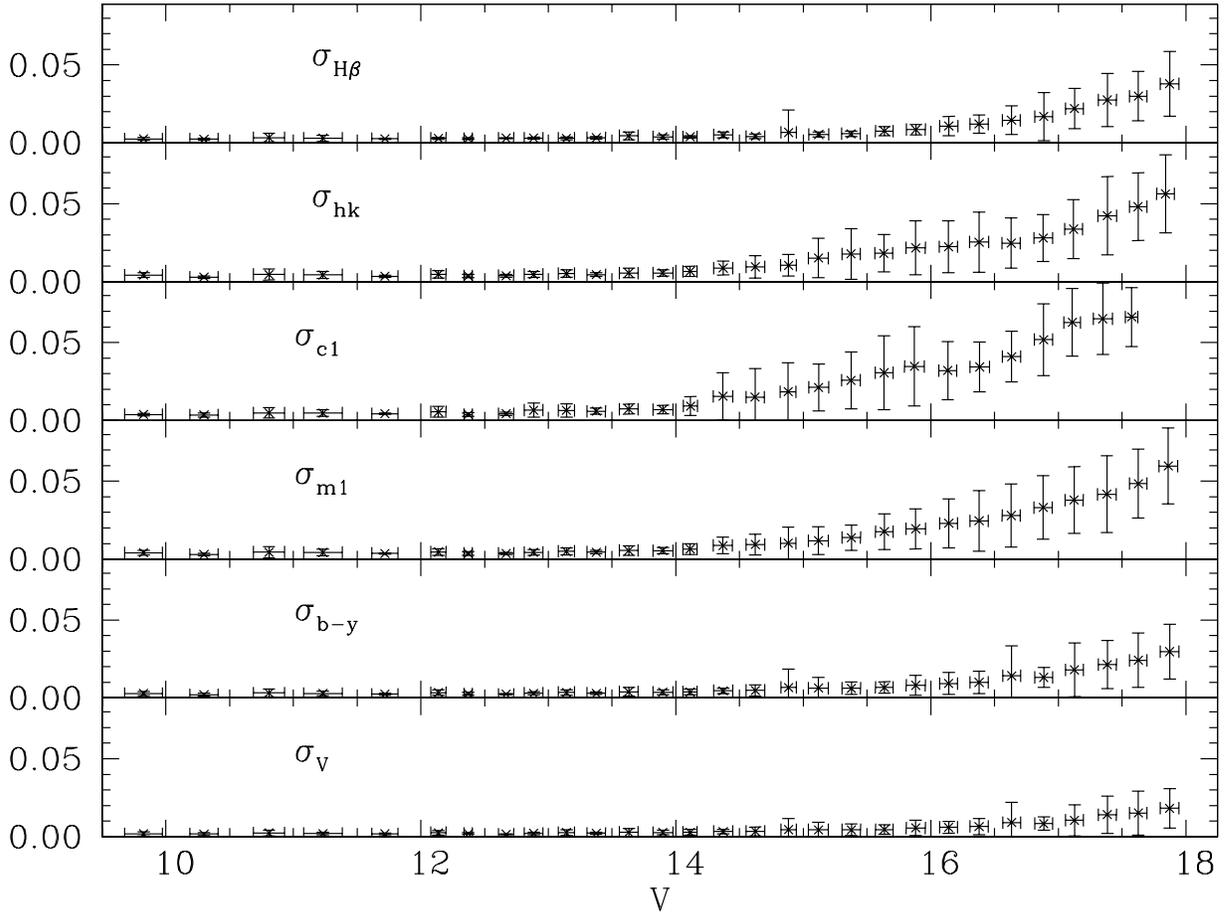}
\caption{Standard errors of the mean as a function of $V$ for the magnitude and indices of Table 5.}
\end{figure}

\clearpage
\begin{figure}
\epsscale{.80}
\centering
\includegraphics[width=\columnwidth]{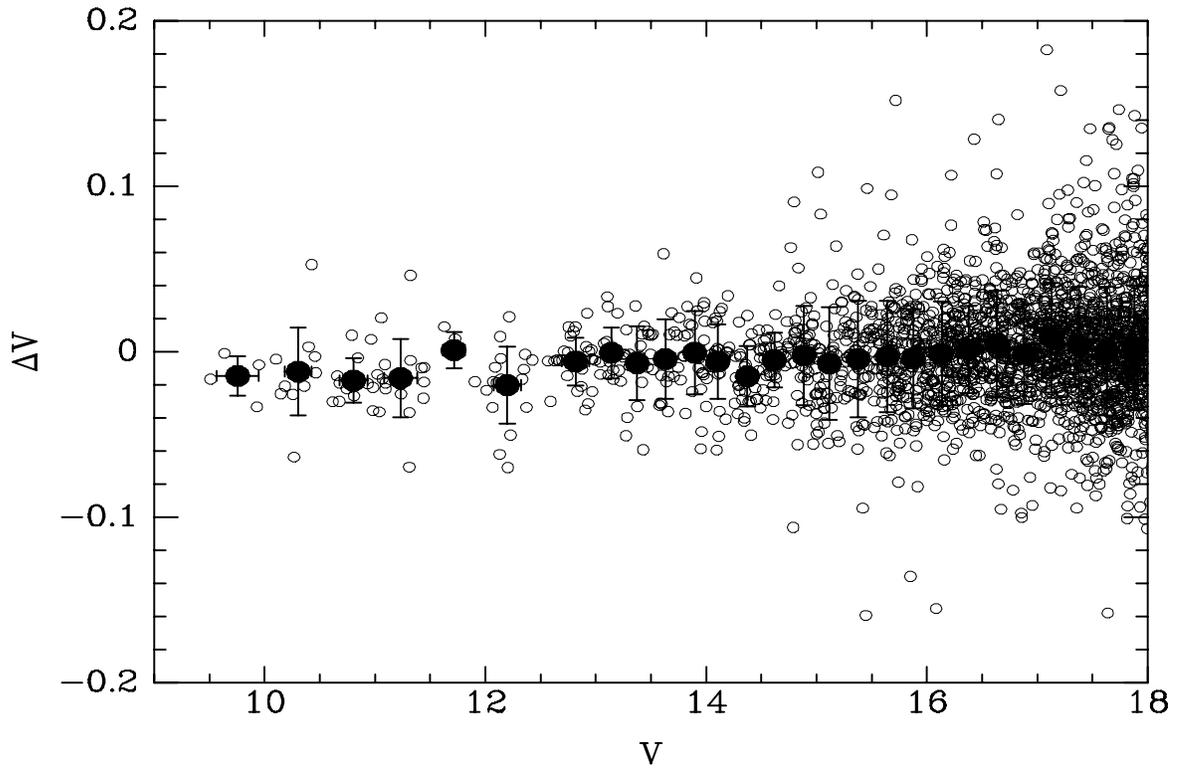}
\caption{Residuals, in the sense (Table 3 - Table 5), between the $V$ mag as a function of $V$.
Filled circles represent the average values within a magnitude bin.}
\end{figure}

\clearpage
\begin{figure}
\epsscale{.80}
\centering
\includegraphics[width=\columnwidth]{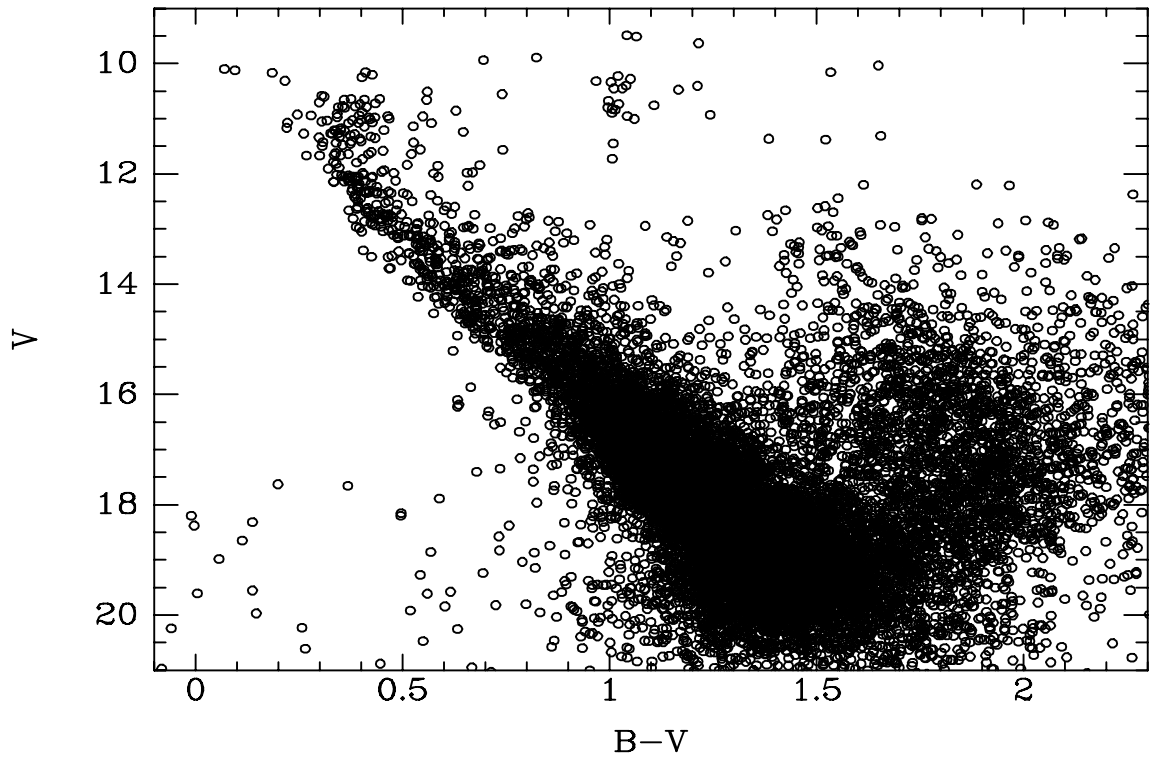}
\caption{The CMD for all stars within the CCD fields, including all stars in Table 3.}
\end{figure}

\clearpage
\begin{figure}
\epsscale{.80}
\centering
\includegraphics[width=\columnwidth]{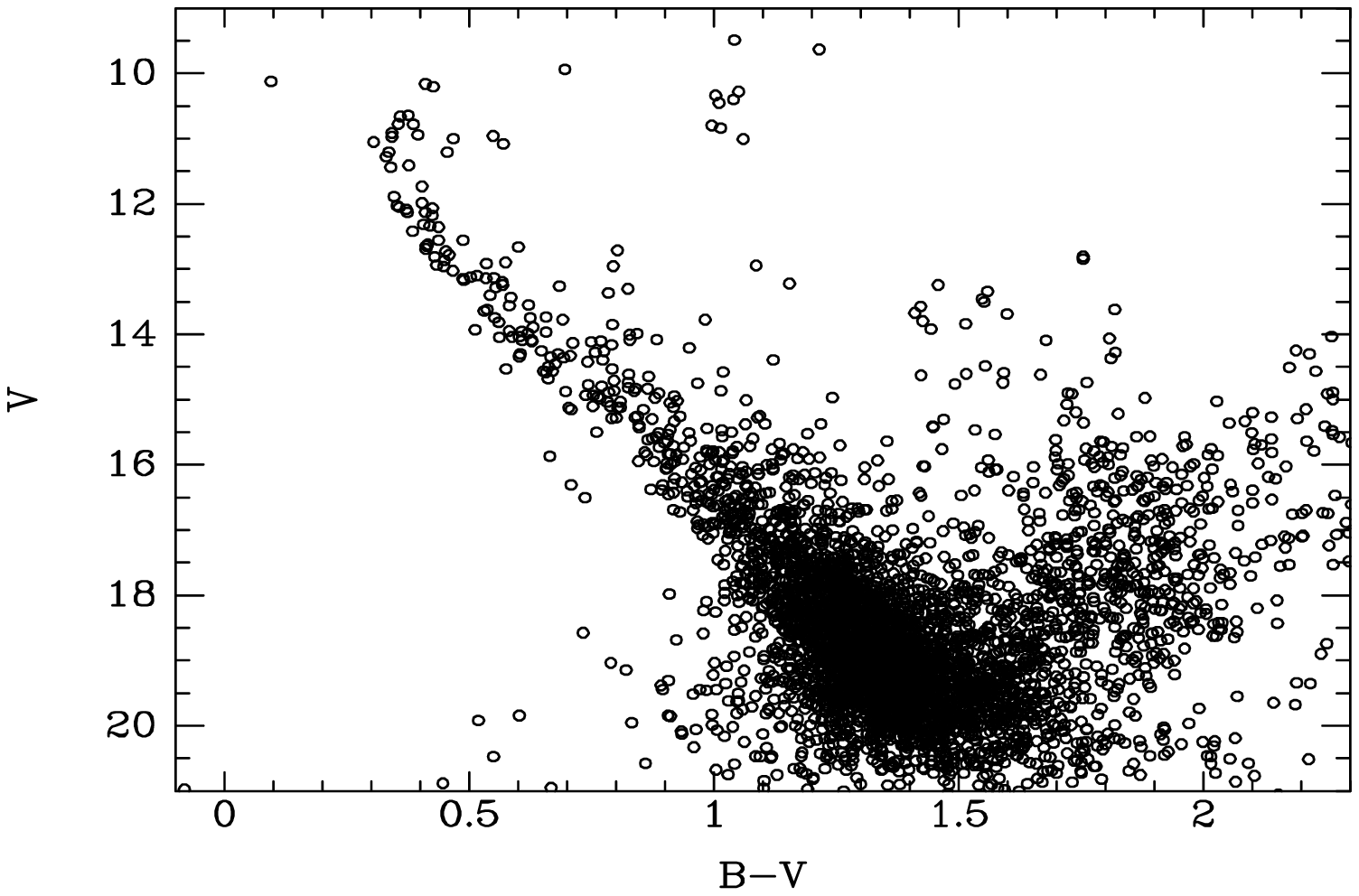}
\caption{Same as Fig. 7 for all stars within a square $13.5^{\prime}$ on a side centered on the cluster.}
\end{figure}

\clearpage
\begin{figure}
\epsscale{.80}
\centering
\includegraphics[width=\columnwidth]{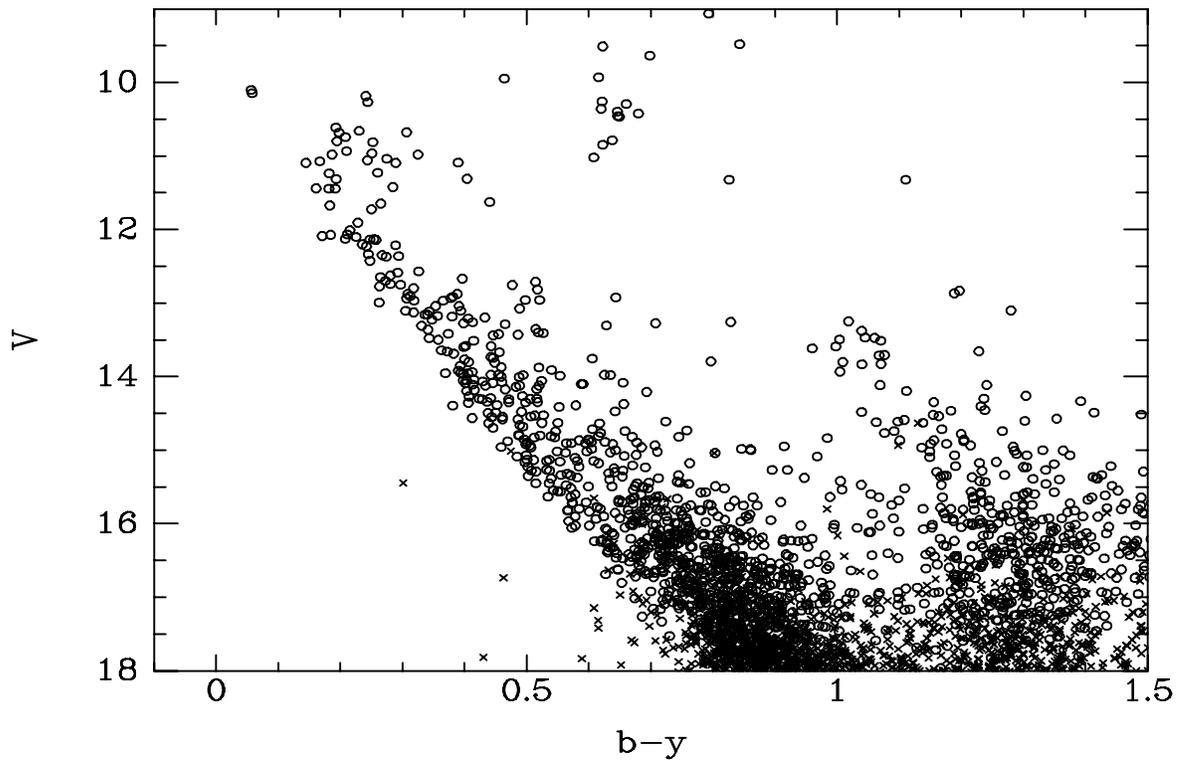}
\caption{The CMD for all stars in Table 5 with errors in $b-y$ no greater than 0.015 mag (open circles) and
no larger than 0.15 mag (crosses).}
\end{figure}

\clearpage
\begin{figure}
\epsscale{.80}
\centering
\includegraphics[width=\columnwidth]{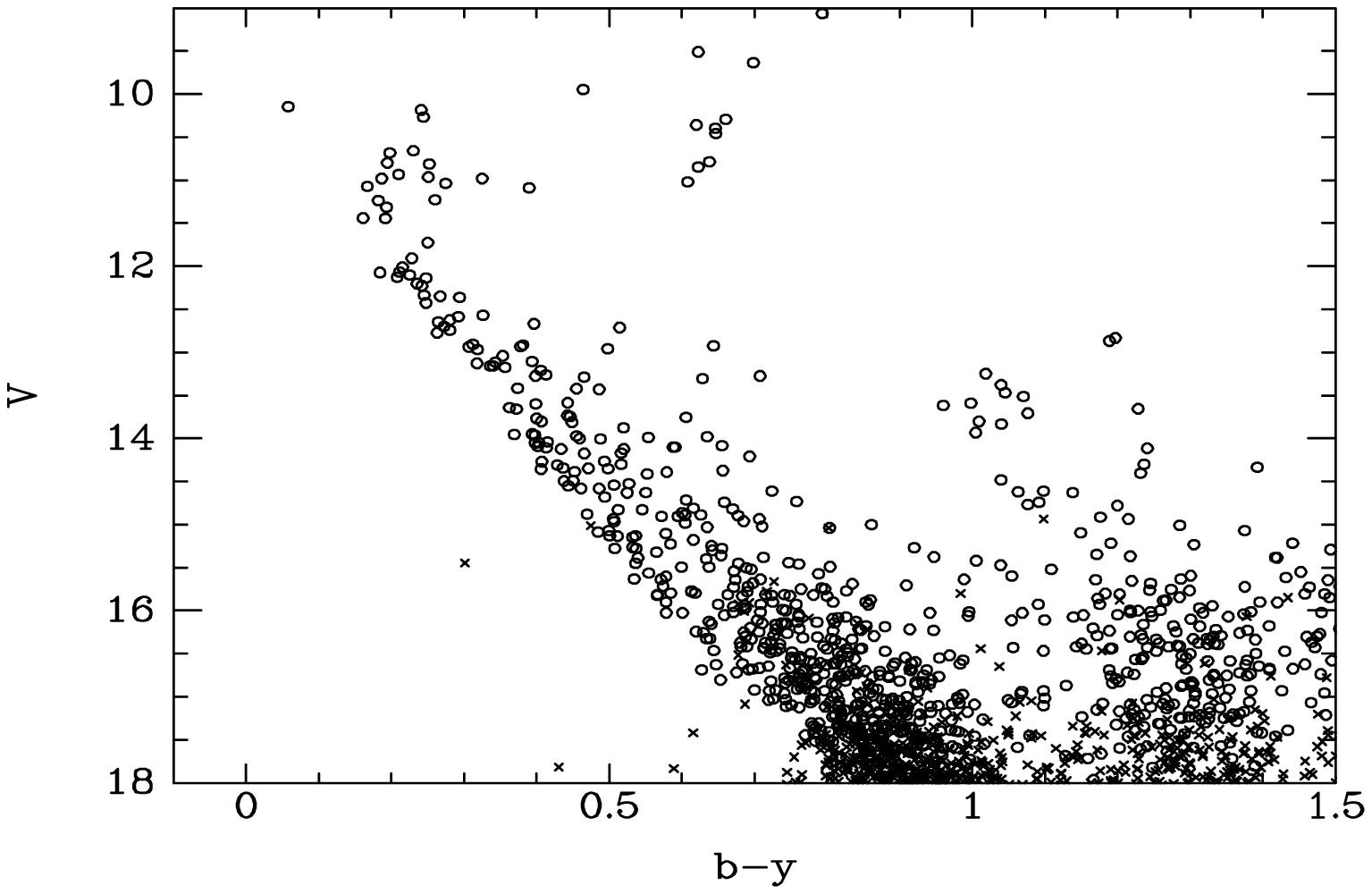}
\caption{Same as Fig. 9 for stars within a square $13.5^{\prime}$ on a side centered on the cluster.}
\end{figure}

\clearpage
\begin{figure}
\epsscale{.80}
\centering
\includegraphics[width=\columnwidth]{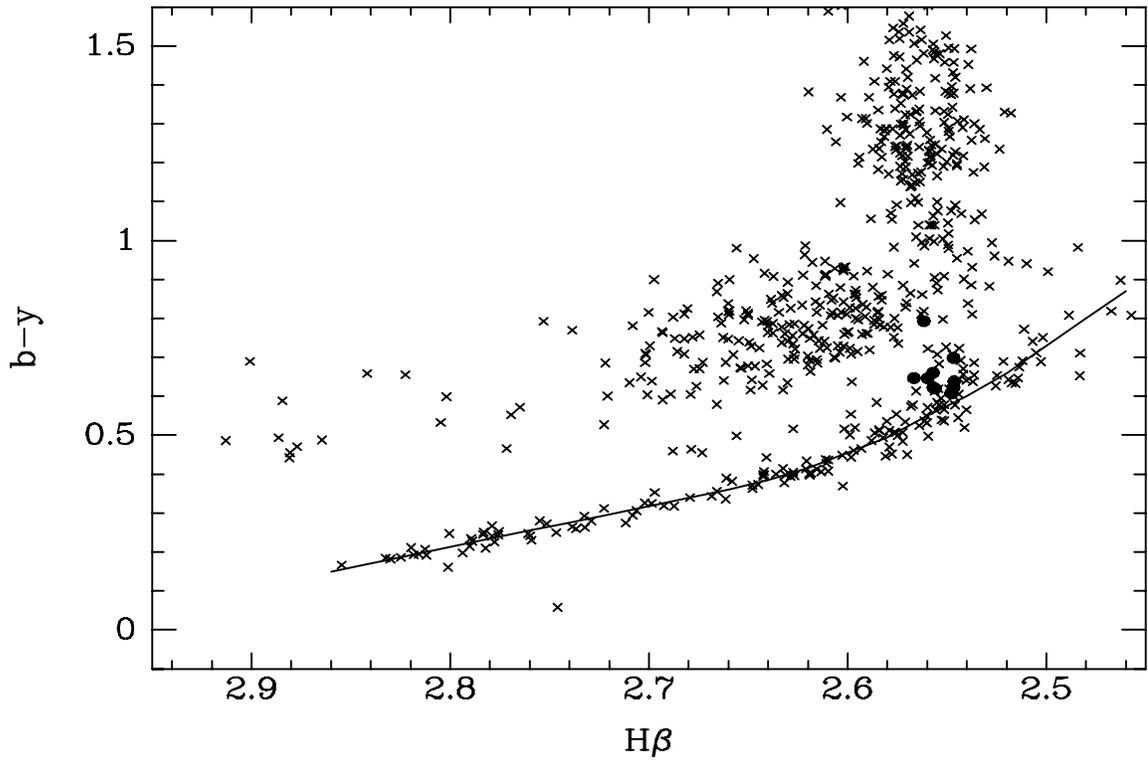}
\caption{Two-color trend for stars in Fig. 10 with at least 2 observations each in H$\beta$ wide and narrow and an sem 
below 0.015 mag in $b-y$ and 0.012 mag for the H$\beta$ index. Filled circles are the
10 giants brighter than $V$ = 11.05. The solid line is the mean relation defined by probable cluster members.}
\end{figure}

\clearpage
\begin{figure}
\epsscale{.80}
\centering
\includegraphics[width=\columnwidth]{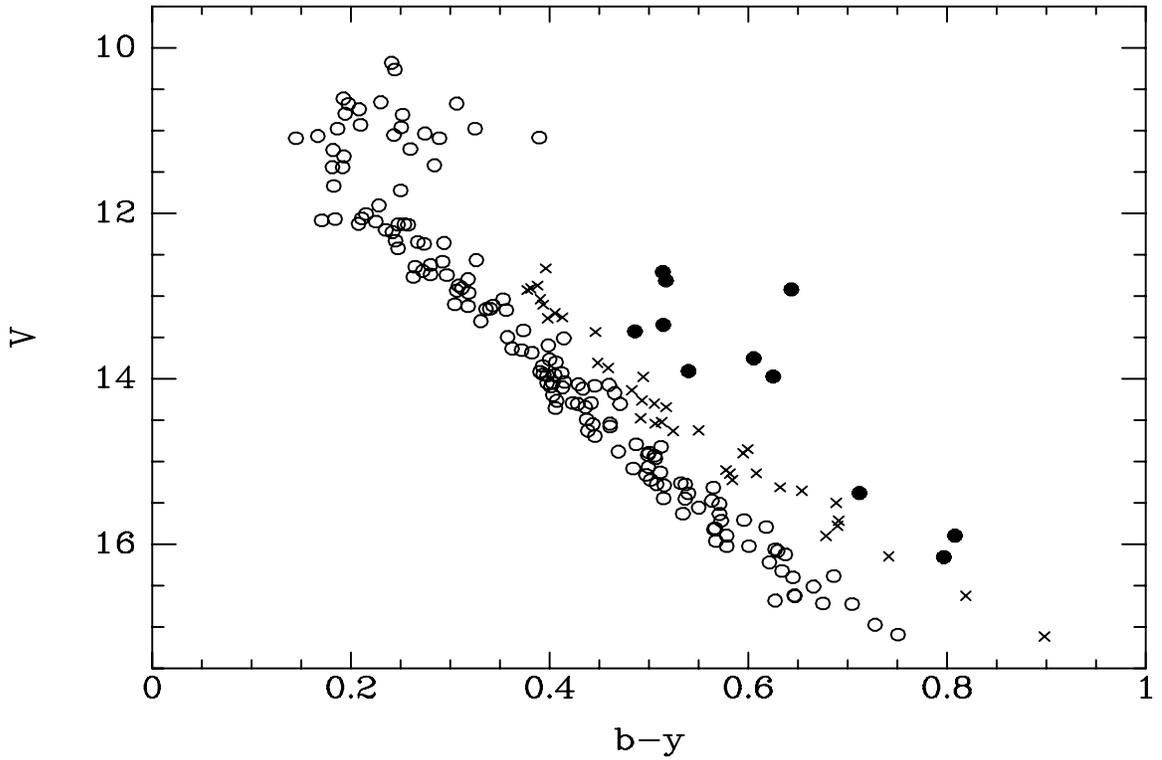}
\caption{$V, b-y$ CMD for stars in Table 5, excluding giants, that lie within 0.04 mag of the mean relation in Fig. 11.  Stars
must be brighter than $V$ = 17.5 with at least 2 observations each in $b, y$, $\beta$ narrow and $\beta$ wide and have errors in $b-y$ 
and H$\beta$ below 0.015 mag. Open circles are probable single-star members, crosses are probable binaries, and
filled circles are likely foreground non-member dwarfs.}
\end{figure}

\clearpage
\begin{figure}
\epsscale{.80}
\centering
\includegraphics[width=\columnwidth]{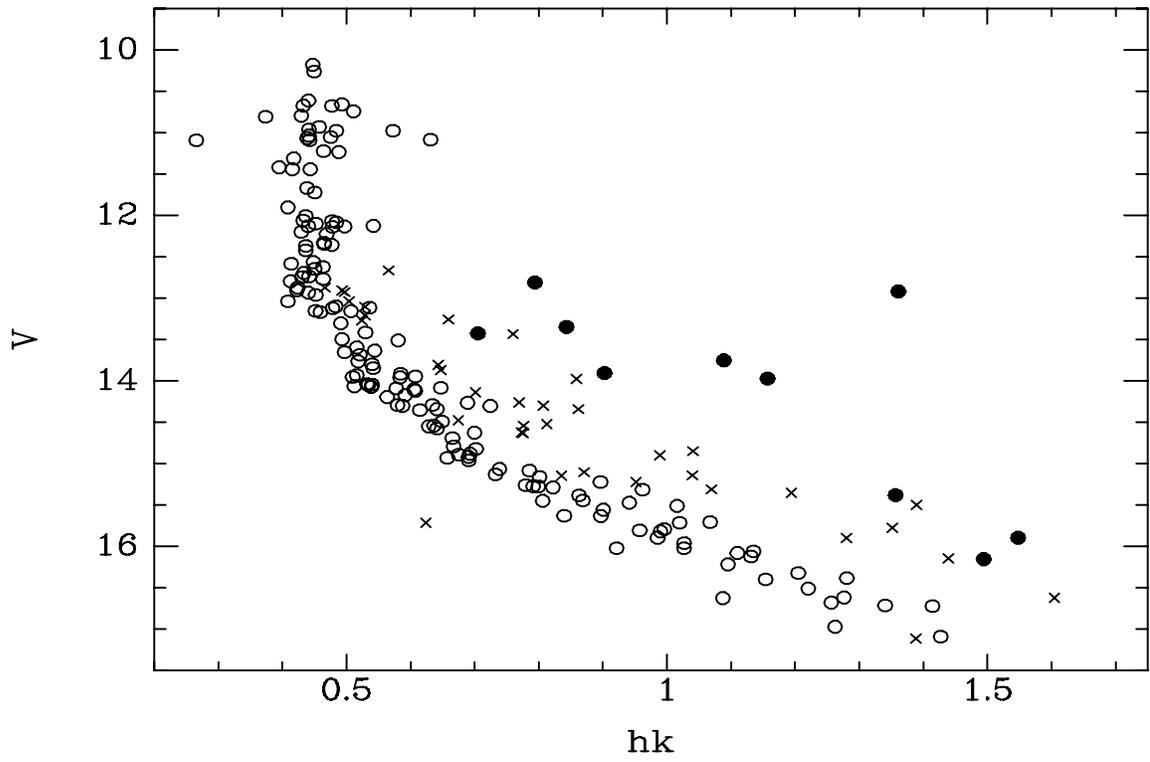}
\caption{Same as Fig. 12 for $V, hk$.}
\end{figure}

\clearpage
\begin{figure}
\epsscale{.80}
\centering
\includegraphics[width=\columnwidth]{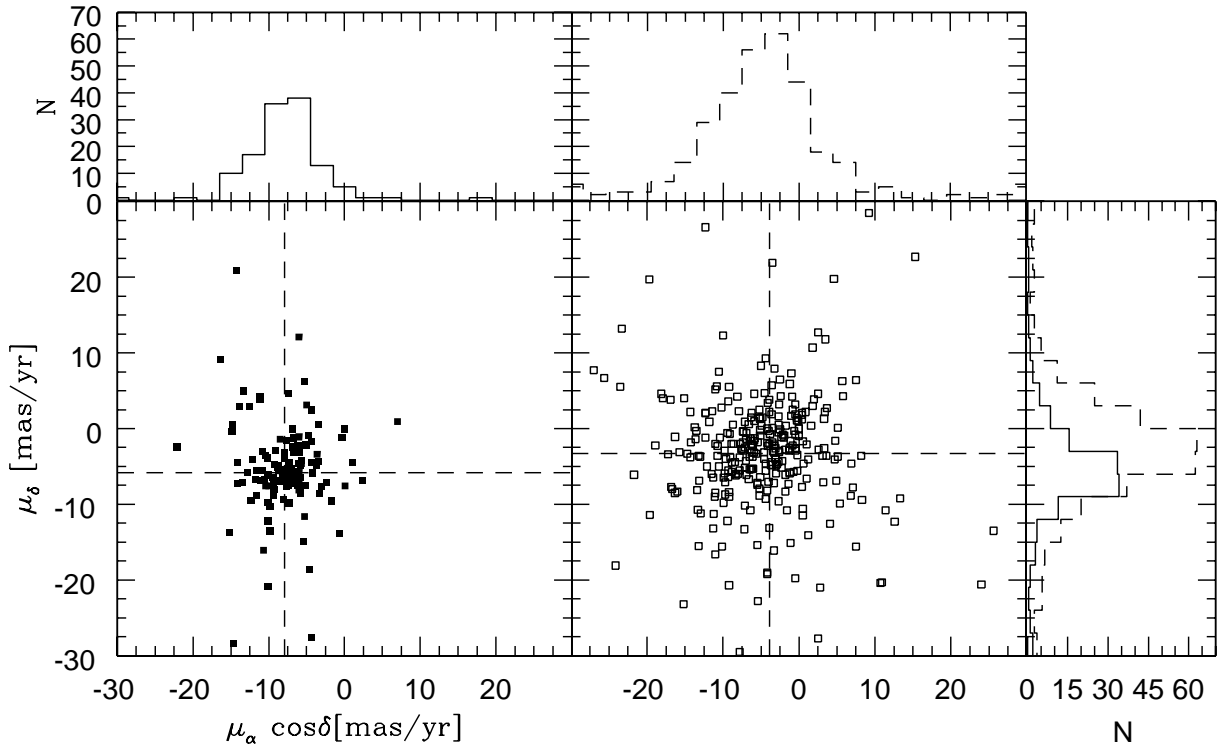}
\caption{UCAC3 proper-motion distributions for photometrically identified members of NGC 5822 (black lines)
and non-members (blue lines). Dashed lines mark the mean motions in each coordinate.}
\end{figure}

\clearpage
\begin{figure}
\epsscale{.80}
\centering
\includegraphics[width=\columnwidth]{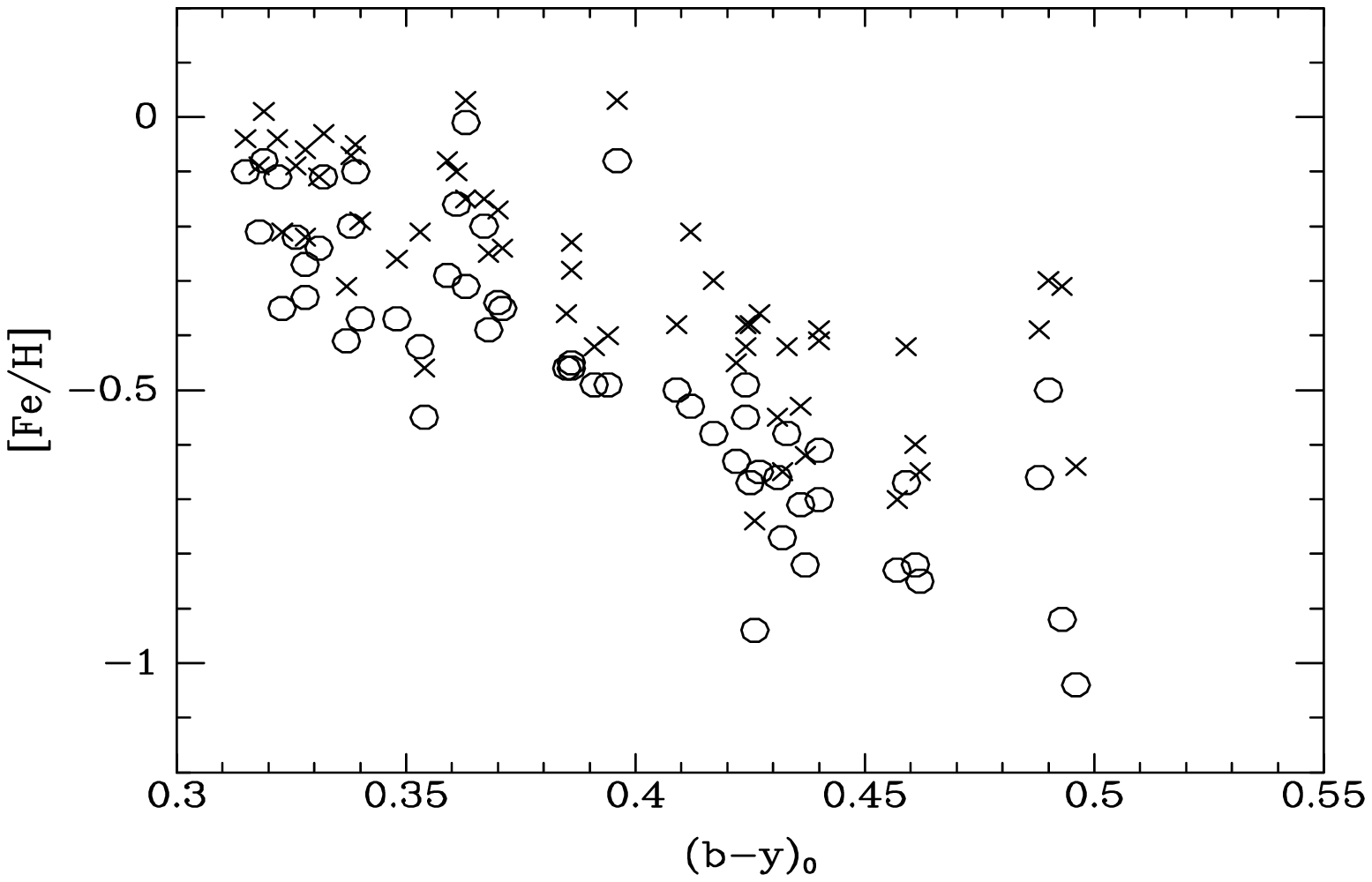}
\caption{[Fe/H] determined using the revised relations of \citet{cas11} (circles) and the H$\beta$-based relations of
\citet{tvat07} (crosses).}
\end{figure}

\clearpage
\begin{figure}
\epsscale{.80}
\centering
\includegraphics[width=\columnwidth]{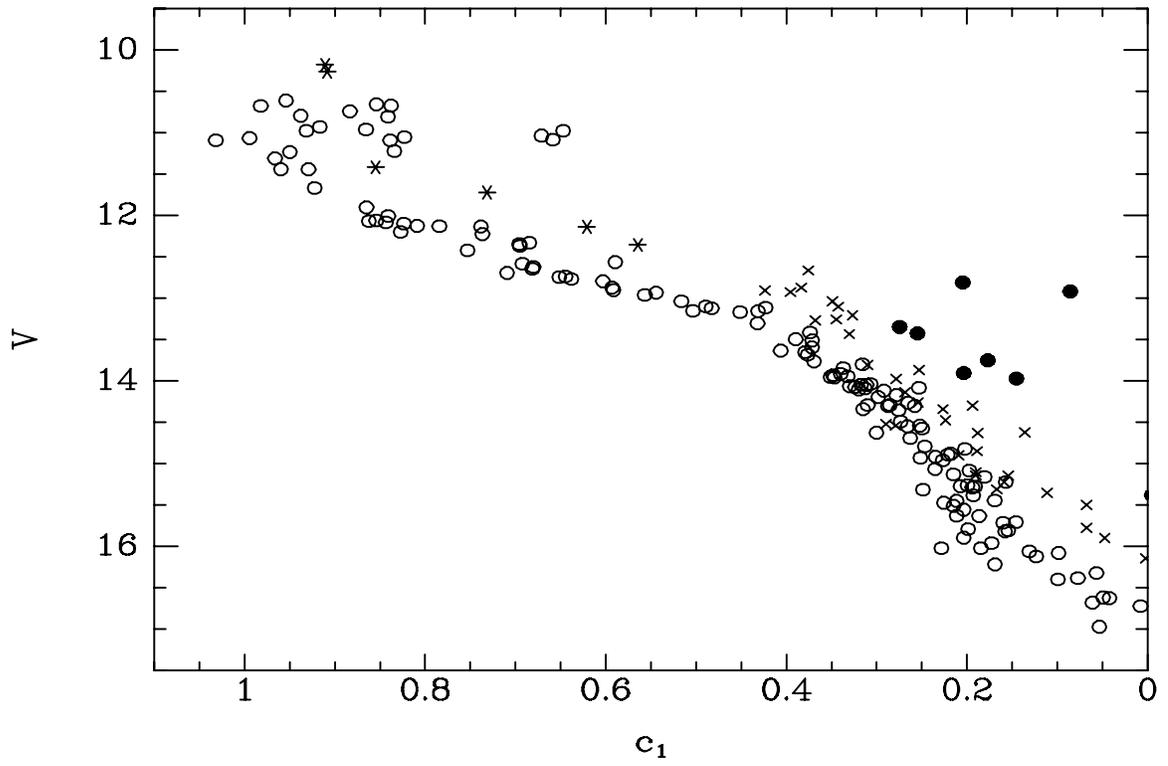}
\caption{Same as Fig. 12 for $V, c_1$. Starred points are newly identified probable binaries.}
\end{figure}

\clearpage
\begin{figure}
\epsscale{.80}
\centering
\includegraphics[width=\columnwidth]{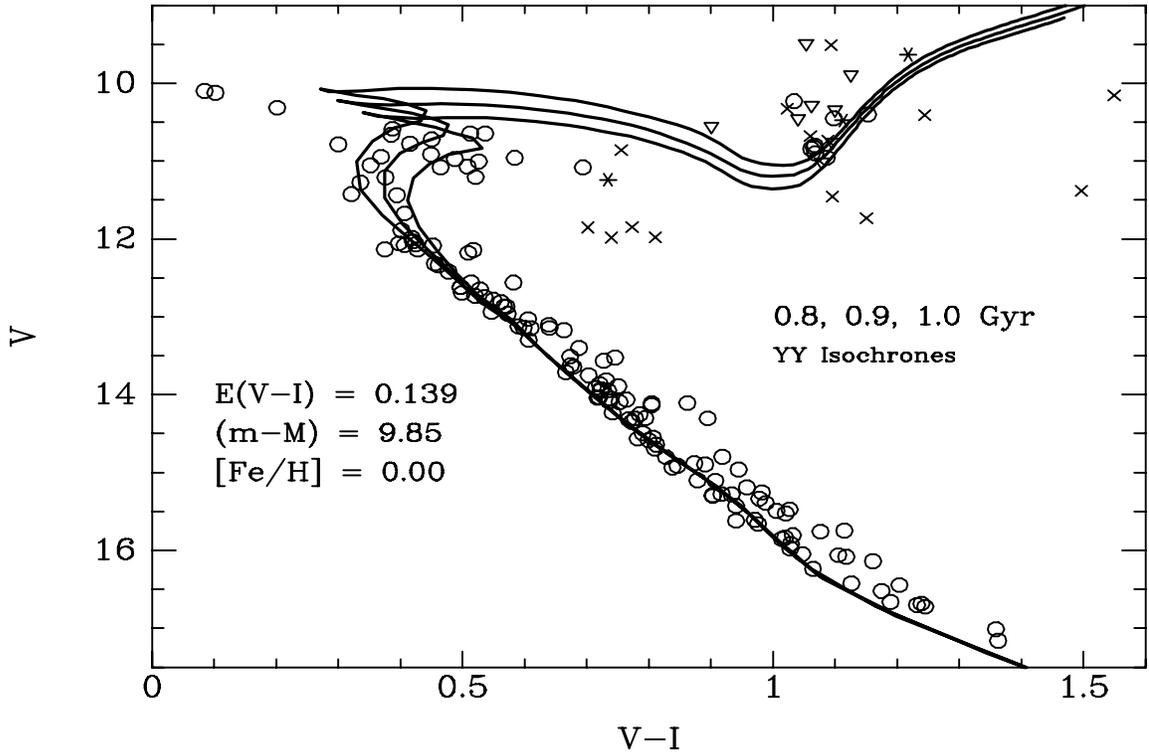}
\caption{The $V, V-I$ CMD compared to $Y^2$ isochrones with [Fe/H] = 0.0, adjusted for E$(V-I)$ = 0.139 and (m-M) = 9.85. Isochrones
have ages of 0.8, 0.9, and 1.0 Gyr. For $V-I < 0.7$, open circles are probable single-star members from photometry. For $V-I > 0.7$,
open circles are radial-velocity, single-star members, open triangles are radial-velocity, binary members, starred points are potential
photometric members, and crosses are stars with only $UVBI$ data.}
\end{figure}

\clearpage
\begin{figure}
\epsscale{.80}
\centering
\includegraphics[width=\columnwidth]{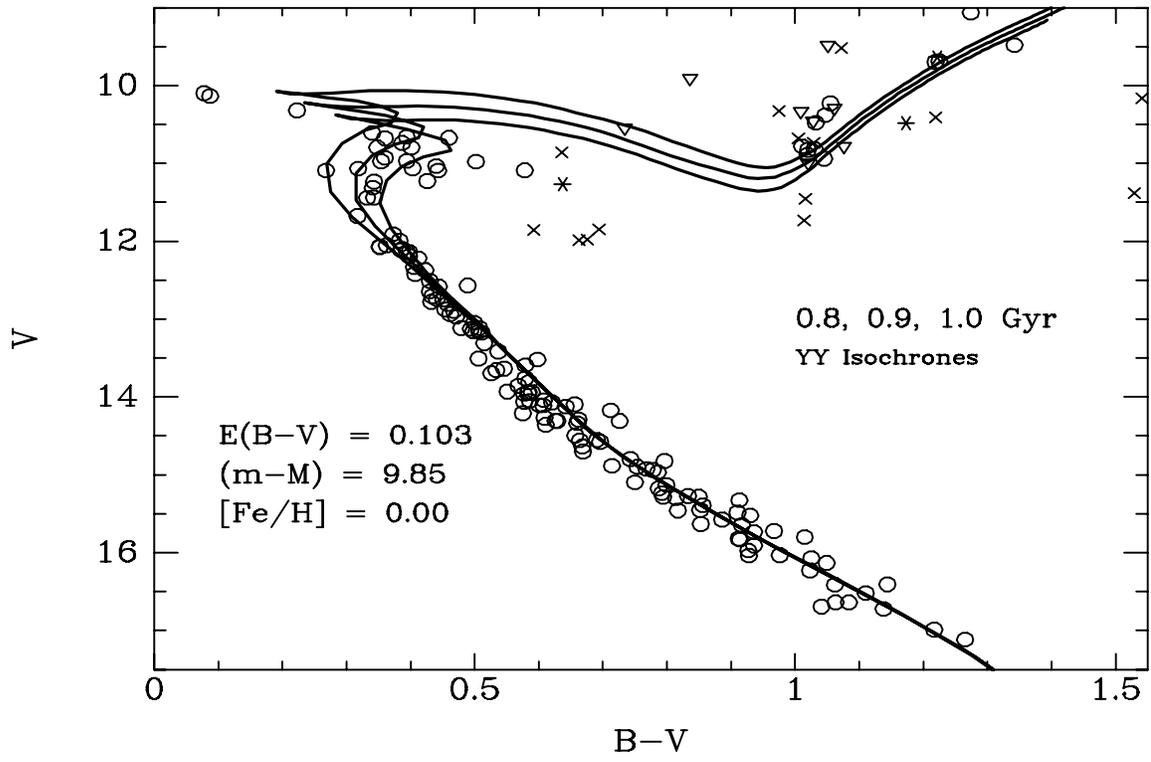}
\caption{Same as Fig. 17 for $E(B-V)$ = 0.103 and $(m-M)$ = 9.85.}
\end{figure}

\clearpage
\begin{figure}
\epsscale{.80}
\centering
\includegraphics[width=\columnwidth]{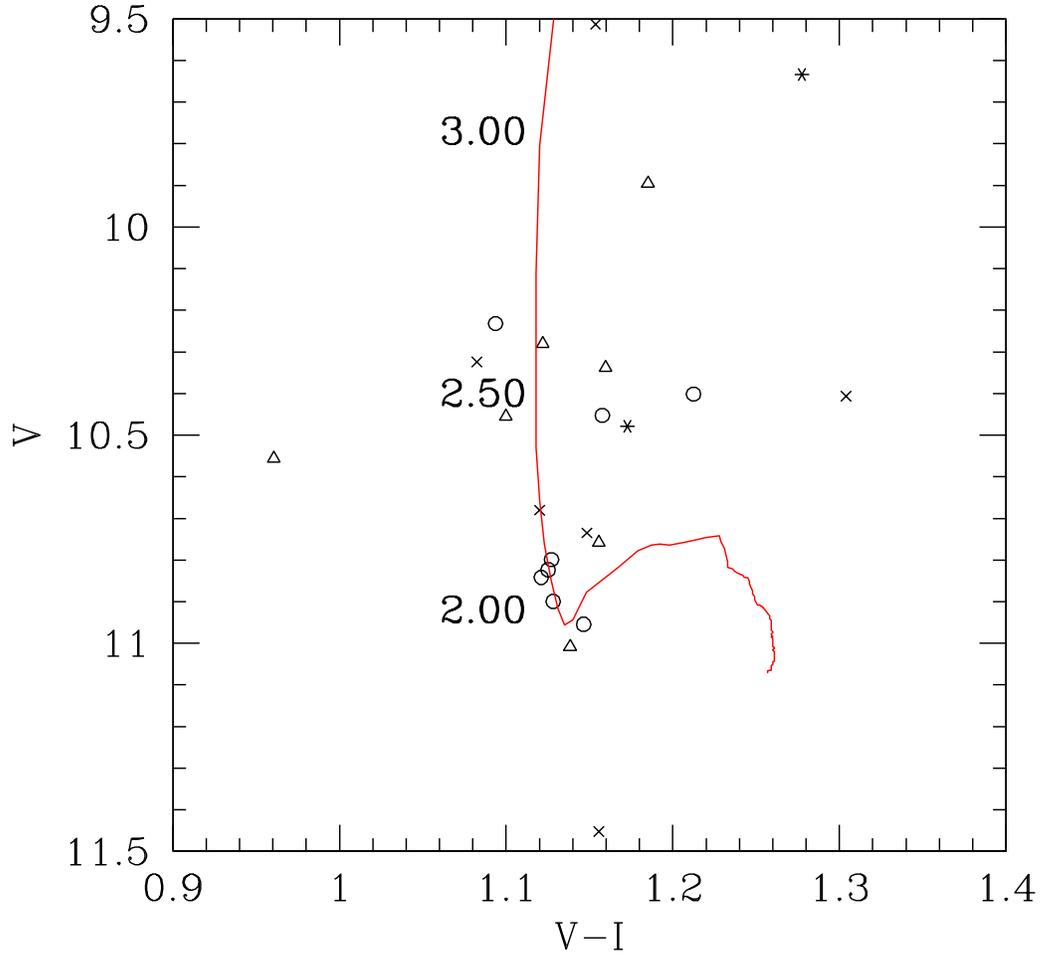}
\caption{The $V, V-I$ CMD of the red giant region compared to core-He-burning models of \citet{gir01}, adjusted to the
distance and reddening of NGC 5822. Initial masses in solar units identify the limiting (faintest) location for evolved stars. 
Symbols have the same meaning as in Fig. 17.}
\end{figure}

\clearpage
\begin{figure}
\epsscale{.80}
\centering
\includegraphics[width=\columnwidth]{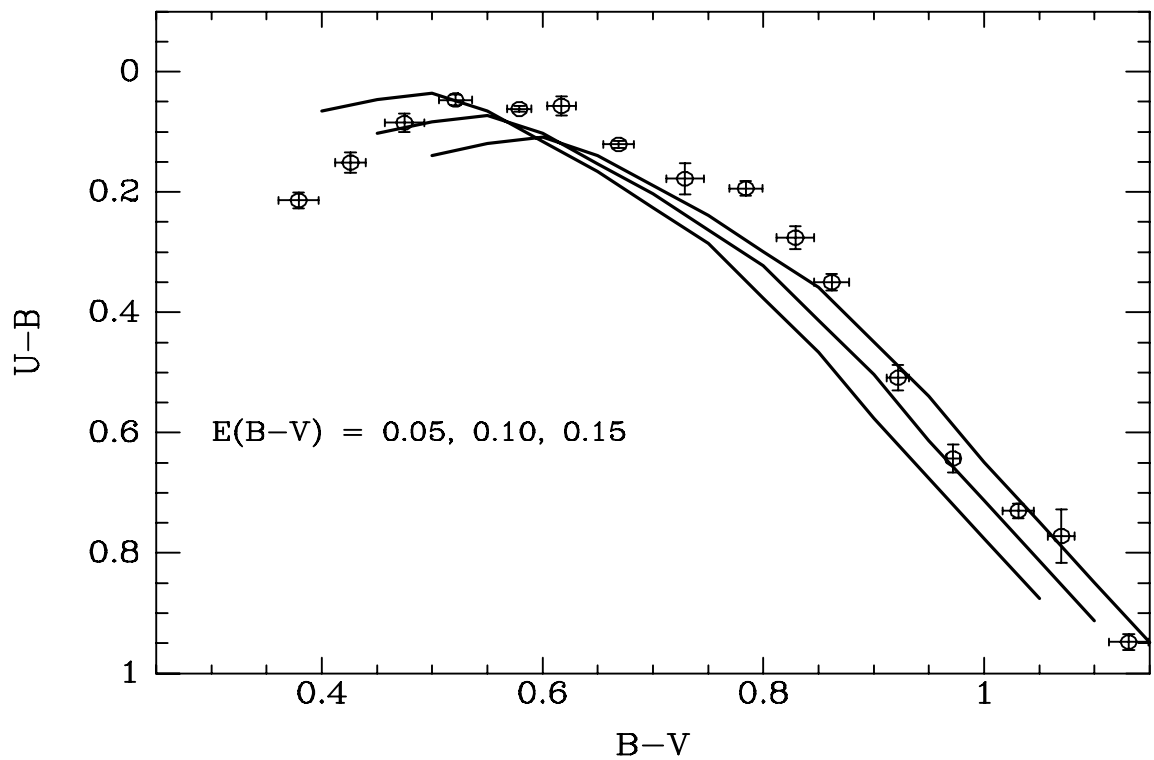}
\caption{Mean two-color data for probable single-star main sequence members of NGC 5822 fainter than $V$ = 12.0.
Error bars in $B-V$ illustrate the dispersion with the binned data while the error bars in $U-B$ are the standard
errors of the mean. Solid lines are the unevolved Hyades relation adjusted for $E(B-V)$ = 0.05, 0.10, and 0.15,
respectively.}
\end{figure}

\newpage
\begin{deluxetable}{ccccc}
\thispagestyle{empty}
\tablecolumns{5}
\tabletypesize\scriptsize
\tablewidth{0pc}
\tablenum{1}
\tablecaption{Basic Cluster Information}
\tablehead{
\colhead{Source and Method} & 
\colhead{E$(B-V)$} & 
\colhead{[Fe/H]} & 
\colhead{$(m-M)$} & 
\colhead{Age(Gyr)} }
\startdata
\cutinhead{Photometry \& Moderate-Dispersion Spectroscopy}
Stetson (1981)  & 0.14 & \nodata & 9.8 & \nodata \\
$uvby$H$\beta$  & DW (21)  & \nodata & DW (12)& \nodata \\
Twarog et al. (1993)  & 0.15 & -0.15 & 9.85 & 1.2--2.0 \\
$UBV$                &  DW (21), RG (17) & DW (21), RG (13) & MSF & ISO \\
                      & 0.14 & -0.11 & & \\
$DDO$                 & RG (16) & RG (15) &   & \\
Twarog et al. (1997)  & 0.15 & -0.02 & 10.0 & \nodata \\
                      &      & $DDO$, RG (14) &  MSF & \nodata \\
                & \nodata  & -0.07       & \nodata & \nodata \\
                & \nodata   & MDS, RG (3)    & \nodata & \nodata \\
Friel et al. (2002) & 0.15 & -0.21 & \nodata & 1.3 \\
                      & \nodata & MDS (3)  & \nodata & MAI \\
This Study      & 0.10 & -0.02 & 9.85 & 0.9 \\
$uvbyCa$H$\beta$ & DW (109) & DW (61) & \nodata & \nodata  \\ 
\cutinhead{High-Dispersion Spectroscopy}
Luck (1994) & \nodata & 0.06 & \nodata & \nodata \\
            &  \nodata & RG (2) & \nodata & \nodata \\
Smiljanic et al. (2009) & \nodata & 0.04 & \nodata & \nodata \\
            &  \nodata & RG (5) & \nodata & \nodata \\
Santos et al. (2009)  & \nodata & 0.05, 0.12\tablenotemark{a} (3) & \nodata & \nodata \\
                      &  \nodata & RG & \nodata & \nodata \\
Pace et al. (2010) & 0.1 & 0.05 & 9.9  & 1.0 \\
                &  T-BV,DW & DW (2) & MSF & ISO \\
                & \nodata & 0.15 & \nodata & \nodata \\
                &  \nodata & RG (3) & \nodata & \nodata \\
\cutinhead{Coordinates and Motion}
$\alpha, \delta\ (2000)$  & 15:04:24.0 & -54:24:05 & &  \\
$l,b$                  & 321.5772 & +3.5851 &  &  \\
Mermilliod et al. (2008) &  &$V_{rad} = -29.31 \pm 0.18$ (20) &  \\
\enddata
\tablenotetext{a}{Different abundances result from use of two different linelists.}
\tablecomments{Distance determinations referred to in the table include main sequence fitting (MSF) and isochrone comparison (ISO).  MAI indicated an age determination using morphological age index; T-BV refers to use of a color-temperature relation for determination of reddening.  Methods applied particularly to dwarfs (DW) and giants (RG) are noted. The number of stars included in each analysis is noted in parentheses. Reddening estimates without a method designation are adopted values for the respective analysis.}

\end{deluxetable}




\newpage
\begin{deluxetable}{lcccccc}

\thispagestyle{empty}
\tablecolumns{7}
\tabletypesize\small
\tablewidth{0pc}
\tablecaption{$UBVI$ photometric observations of NGC~5822 and standard star fields.}
\tablenum{2}
\tablehead{
\colhead{Target} & \colhead{Date} & \colhead{RA (2000.0)} &
\colhead{DEC (2000.0)} & \colhead{Filter} & \colhead{Exposures(sec)} &
\colhead{Airmass}} 

\startdata
SA~98        & 2009 Mar 19     & 06:52:04.4 & -00:19:36 &\textit{U} & 2x20,2x150,2x400     &1.17$-$2.35\cr
             &                 &            &           &\textit{B} & 2x20,2x100,2x200     &1.16$-$2.04\cr
             &                 &            &           &\textit{V} & 2x10,2x60,2x120      &1.15$-$1.80\cr
             &                 &            &           &\textit{I} & 2x10,2x60,2x120      &1.15$-$1.92\cr
NGC~5822 (E) & 2009 Mar 19     & 15:04:20.0 & -54:23:49 &\textit{U} & 30, 200, 2000        &1.28$-$1.30\cr
             &                 &            &           &\textit{B} & 20, 150, 1500        &1.10$-$1.20\cr
             &                 &            &           &\textsl{V} & 10, 100, 900         &1.14$-$1.15\cr
             &                 &            &           &\textsl{I} & 10, 100, 900         &1.10$-$1.10\cr
PG~1047      & 2009 Mar 19     & 10:50:11.3 & -00:01:06 &\textit{U} & 30, 400              &1.46$-$1.48\cr
             &                 &            &           &\textit{B} & 20, 200              &1.42$-$1.44\cr
             &                 &            &           &\textit{V} & 20, 100              &1.37$-$1.38\cr
             &                 &            &           &\textit{I} & 20, 100              &1.40$-$1.41\cr
NGC~5822 (A) & 2010 Mar 11     & 15:05:29.1 & -54:12:44 &\textit{U} & 30, 200, 2000        &1.08$-$1.17\cr
             &                 &            &           &\textit{B} & 20, 150, 1500        &1.18$-$1.20\cr
             &                 &            &           &\textsl{V} & 10, 100, 900         &1.21$-$1.22\cr
             &                 &            &           &\textsl{I} & 10, 100, 900         &1.22$-$1.24\cr
NGC~5822 (B) & 2010 Mar 12     & 15:05:36.3 & -54:31:49 &\textit{U} & 30, 200, 2000        &1.11$-$1.20\cr
             &                 &            &           &\textit{B} & 20, 150, 1500        &1.20$-$1.28\cr
             &                 &            &           &\textsl{V} & 10, 100, 900         &1.28$-$1.31\cr
             &                 &            &           &\textsl{I} & 10, 100, 900         &1.31$-$1.33\cr
NGC~5822 (C) & 2010 Mar 13     & 15:03:11.8 & -54:33:01 &\textit{U} & 30, 200, 2000        &1.10$-$1.30\cr
             &                 &            &           &\textit{B} & 20, 150, 1500        &1.31$-$1.40\cr
             &                 &            &           &\textsl{V} & 10, 100, 900         &1.42$-$1.45\cr
             &                 &            &           &\textsl{I} & 10, 100, 900         &1.45$-$1.45\cr
NGC~5822 (D) & 2010 Mar 14     & 15:03:11.1 & -54:11:59 &\textit{U} & 30, 200, 2000        &1.06$-$1.12\cr
             &                 &            &           &\textit{B} & 20, 150, 1500        &1.12$-$1.19\cr
             &                 &            &           &\textsl{V} & 10, 100, 900         &1.21$-$1.25\cr
             &                 &            &           &\textsl{I} & 10, 100, 900         &1.25$-$1.28\cr
\enddata
\end{deluxetable}

\newpage
\begin{deluxetable}{rrrrrrrrrr}
\pagestyle{empty}
\tablecolumns{10}
\tablenum{3}
\tabletypesize\small
\tablewidth{0pc}
\tablecaption{Broad-Band Photometry of Stars in the Field of NGC 5822}
\tablehead{
\colhead{$\alpha(2000)$} & \colhead{$\delta(2000)$} & \colhead{$V$} & \colhead{$\sigma_V$} &
\colhead{$B-V$} & \colhead{$\sigma_{BV}$} & 
\colhead{$V-I$} & \colhead{$\sigma_{VI}$} &
\colhead{$U-B$} & \colhead{$\sigma_{UB}$} }
\startdata 
     225.9222  &  -54.3758 &  9.493 &  0.018 &  1.042 &  0.024 &  1.053  & 0.029  & 0.709 &  0.034 \\
     226.4440  &  -54.7140 &  9.513 &  0.042 &  1.065 &  0.024 &  1.094  & 0.059  & 0.624 &  0.034 \\
     226.2494  &  -54.2987 &  9.634 &  0.017 &  1.215 &  0.025 &  1.217  & 0.032  & 1.106 &  0.040 \\
     226.1497  &  -54.5566 &  9.896 &  0.019 &  0.823 &  0.023 &  1.125  & 0.058  & 0.535 &  0.033 \\
     226.0530  &  -54.4169 &  9.942 &  0.015 &  0.695 &  0.022 &  0.870  & 0.024  & 0.486 &  0.032 \\
     226.3524  &  -54.3832 & 10.037 &  0.019 &  1.649 &  0.030 &  1.723  & 0.027  & 2.061 &  0.057 \\
     225.9789  &  -54.5523 & 10.098 &  0.017 &  0.070 &  0.024 &  0.084  & 0.028  &-0.083 &  0.033 \\
     226.2603  &  -54.4403 & 10.122 &  0.015 &  0.095 &  0.020 &  0.102  & 0.021  &-0.290 &  0.031 \\
     226.4561  &  -54.1401 & 10.159 &  0.023 &  1.534 &  0.030 &  1.549  & 0.072  & 1.229 &  0.042 \\
     226.2061  &  -54.5067 & 10.163 &  0.015 &  0.410 &  0.021 &  0.478  & 0.021  & 0.233 &  0.031 \\
           & & & & & & & & & \\
     226.5713  &  -54.5963 & 10.170 &  0.019 &  0.185 &  0.023 &  0.163  & 0.025  & 0.026 &  0.032 \\
     226.0712  &  -54.3941 & 10.202 &  0.014 &  0.427 &  0.020 &  0.475  & 0.021  & 0.302 &  0.031 \\
     225.9583  &  -54.2418 & 10.232 &  0.016 &  1.020 &  0.024 &  1.034  & 0.026  & 0.703 &  0.034 \\
     226.6095  &  -54.1983 & 10.250 &  0.025 &  0.402 &  0.021 &  0.406  & 0.036  & 0.246 &  0.032 \\
     226.0591  &  -54.4298 & 10.281 &  0.016 &  1.050 &  0.024 &  1.062  & 0.024  & 0.771 &  0.035 \\
     226.4627  &  -54.4184 & 10.315 &  0.015 &  0.216 &  0.020 &  0.201  & 0.022  & 0.148 &  0.031 \\
     226.5692  &  -54.2922 & 10.324 &  0.028 &  0.968 &  0.024 &  1.023  & 0.036  & 0.542 &  0.033 \\
     226.1794  &  -54.3059 & 10.338 &  0.016 &  1.003 &  0.024 &  1.100  & 0.024  & 0.608 &  0.033 \\
     226.1643  &  -54.3512 & 10.401 &  0.016 &  1.040 &  0.024 &  1.153  & 0.023  & 0.775 &  0.035 \\
     226.6916  &  -54.2093 & 10.407 &  0.017 &  1.212 &  0.026 &  1.244  & 0.035  & 1.004 &  0.038 \\
\enddata
\end{deluxetable}

\newpage
\begin{deluxetable}{lccccccccc}
\pagestyle{empty}
\tablecolumns{9}
\tablenum{4}
\tabletypesize\small
\tablewidth{0pc}
\tablecaption{Characterization of Calibration Equations}
\tablehead{
\colhead{} & \colhead{$V$} & \colhead{$b-y$} & \colhead{$m_1 (bd)$} &
\colhead{$m_1 (rg)$} & \colhead{$c_1 (bd)$} & \colhead{$c_1 (rg)$} &
\colhead{$hk$} & \colhead{H$\beta$}}
\startdata
Number of photometric nights used& 4 & 4 & 2 &2  & 2 & 2 & 4 & 2  \cr
Calibration equation slope $\alpha$ & 1.000 & 1.025 & 0.940 & 1.050 & 1.020 & 0.900 & 1.070& 1.170 \cr
Color term $\gamma$ & 0.05 & \nodata  &  0.11& -0.09 & -0.14 & 0.25 & \nodata  & \nodata \cr
Maximum standard deviation for & 0.022 & 0.019 & 0.015 & 0.031 & 0.018 & 0.053 & 0.034 & 0.021 \cr
\phm{word}calibration equation for one night & & & &  & & & & \cr
Typical contribution to zeropoint ($\beta$)  & 0.002 & 0.002 & 0.003 & \nodata & 0.004 & \nodata & 0.003 & 0.002 \cr
\phm{word}s.e.m. from aperture correction & & & & & & & &  \cr
Combined s.e.m. for final & 0.004 & 0.005 &  0.016 & 0.020 & 0.017 & 0.030 & 0.007 & 0.005 \cr
\phm{word}calibration equation & & & & & & & &  \cr
Number of photoelectric & 7-17 & 5-15 & 3-4 & 11-14 & 3-4 & 11-14 & 7-17 & 6-15 \cr
\phm{word}standards in index calibration & & & & & & & & \cr
\enddata
\tablecomments{Calibration equations for index $x$ are of the form $x_{std} = \alpha\  x_{instr} + \gamma (b-y)_{instr} + \beta$. 
Classes of stars include warm dwarfs {\it bd} and cool giants {\it rg}.}
\end{deluxetable}

\newpage
\begin{deluxetable}{rrrrrrrrrrrrrrrrrrrrr}
\pagestyle{empty}
\rotate
\tablecolumns{21}
\tablenum{5}
\tabletypesize\scriptsize
\tablewidth{0pc}
\tablecaption{$uvbyCa$H$\beta$ Photometry of Stars in the Field of NGC 5822}
\tablehead{
\colhead{$\alpha(2000)$} & \colhead{$\delta(2000)$} & \colhead{$V$} & 
\colhead{$b-y$} & \colhead{$m_1$} & \colhead{$c_1$} & \colhead{$hk$} & \colhead{H$\beta$} &
\colhead{$\sigma_V$} & \colhead{$\sigma_{by}$} & \colhead{$\sigma_{m1}$} & 
\colhead{$\sigma_{c1}$} & \colhead{$\sigma_{hk}$} & \colhead{$\sigma_{\beta}$} & 
\colhead{$N_y$} & \colhead{$N_b$} & \colhead{$N_v$} & \colhead{$N_u$} & 
\colhead{$N_{Ca}$} & \colhead{$N_n$} & \colhead{$N_w$} }
\startdata 
   226.0096 & -54.3398 &  9.067 &  0.793 &  0.505 &  0.347  & 1.296 &  2.562 &  0.001 &  0.001 &  0.002 &  0.003 &  0.002 &  0.002 &31& 34 &25 &25 &32 &25 &25 \\
   226.1272 & -54.5271 &  9.478 &  0.843 &  0.558 &  0.316  & 1.433 &  2.556 &  0.003 &  0.004 &  0.004 &  0.004 &  0.004 &  0.001 &15& 17 &17 &17 &19 &15 &15 \\
   225.9233 & -54.3762 &  9.510 &  0.623 &  0.396 &  0.355  & 1.048 &  2.547 &  0.002 &  0.003 &  0.004 &  0.004 &  0.003 &  0.004 &25& 24 &25 &25 &28 &15 &17 \\
   226.2482 & -54.2994 &  9.635 &  0.698 &  0.487 &  0.428  & 1.263 &  2.547 &  0.001 &  0.002 &  0.003 &  0.003 &  0.004 &  0.004 &25& 27 &25 &25 &32 &24 &23 \\
   226.1502 & -54.5542 &  9.929 &  0.615 &  0.026 &  0.890  & 0.371 &  2.688 &  0.004 &  0.005 &  0.006 &  0.005 &  0.006 &  0.002 &18& 18 &17 &17 &19 &15 &15 \\
   226.0536 & -54.4164 &  9.950 &  0.464 &  0.178 &  0.774  & 0.604 &  2.679 &  0.001 &  0.002 &  0.003 &  0.003 &  0.003 &  0.002 &36& 37 &26 &26 &35 &24 &24 \\
   225.9813 & -54.5503 & 10.103 &  0.056 &  0.072 &  0.889  & 0.194 &  2.842 &  0.001 &  0.002 &  0.002 &  0.002 &  0.002 &  0.003 &19& 19 &17 &17 &19 &15 &15 \\
   226.2593 & -54.4391 & 10.148 &  0.058 &  0.051 &  0.702  & 0.163 &  2.746 &  0.001 &  0.001 &  0.002 &  0.002 &  0.002 &  0.002 &35& 32 &26 &26 &34 &23 &24 \\
   226.2057 & -54.5048 & 10.184 &  0.241 &  0.141 &  0.911  & 0.447 &  2.779 &  0.004 &  0.004 &  0.005 &  0.004 &  0.005 &  0.003 &19& 19 &17 &17 &19 &15 &15 \\
   225.9581 & -54.2437 & 10.258 &  0.622 &  0.375 &  0.333  & 0.991 &  2.559 &  0.002 &  0.002 &  0.003 &  0.004 &  0.003 &  0.001 &16& 16 &16 &16 &16 &16 &16 \\
            & & & & & & & & & & & & & & & & & & & & \\
   226.0715 & -54.3938 & 10.266 &  0.244 &  0.142 &  0.909  & 0.449 &  2.776 &  0.001 &  0.002 &  0.003 &  0.003 &  0.002 &  0.002 &33& 34 &25 &25 &32 &25 &25 \\
   226.0597 & -54.4291 & 10.292 &  0.660 &  0.331 &  0.454  & 1.051 &  2.557 &  0.002 &  0.002 &  0.004 &  0.004 &  0.003 &  0.003 &36& 37 &26 &26 &35 &23 &24 \\
   226.1785 & -54.3066 & 10.359 &  0.620 &  0.270 &  0.457  & 0.913 &  2.556 &  0.002 &  0.002 &  0.003 &  0.004 &  0.003 &  0.003 &32& 34 &25 &22 &31 &25 &25 \\
   226.1637 & -54.3513 & 10.399 &  0.646 &  0.339 &  0.460  & 1.060 &  2.560 &  0.002 &  0.002 &  0.003 &  0.003 &  0.002 &  0.002 &33& 34 &25 &25 &32 &25 &25 \\
   226.2304 & -54.6147 & 10.426 &  0.680 &  0.449 &  0.411  & 1.194 &  2.565 &  0.001 &  0.001 &  0.003 &  0.007 &  0.003 &  0.003 &19& 19 &17 &17 &15 &15 &15 \\
   226.0719 & -54.4719 & 10.458 &  0.647 &  0.300 &  0.472  & 1.000 &  2.567 &  0.001 &  0.002 &  0.002 &  0.002 &  0.003 &  0.002 &36& 37 &26 &26 &35 &21 &22 \\
   226.1281 & -54.5940 & 10.465 &  0.649 &  0.342 &  0.456  & 1.049 &  2.569 &  0.003 &  0.003 &  0.003 &  0.004 &  0.003 &  0.002 &19& 19 &17 &17 &19 &15 &15 \\
   225.9722 & -54.2369 & 10.616 &  0.193 &  0.149 &  0.954  & 0.441 &  2.820 &  0.001 &  0.003 &  0.004 &  0.003 &  0.004 &  0.001 &16& 16 &16 &16 &16 &16 &16 \\
   225.9115 & -54.3010 & 10.660 &  0.230 &  0.177 &  0.854  & 0.493 &  2.759 &  0.003 &  0.004 &  0.005 &  0.006 &  0.005 &  0.010 &23& 17 &21 &25 &23 &15 &10 \\
   226.1092 & -54.5226 & 10.677 &  0.306 &  0.122 &  0.838  & 0.433 &  2.737 &  0.003 &  0.004 &  0.005 &  0.004 &  0.005 &  0.003 &19& 19 &17 &17 &19 &15 &15 \\
\enddata
\end{deluxetable}

\end{document}